\documentclass{IEEEtran}
\usepackage[latin9]{inputenc}
\usepackage{mathtools}
\usepackage{amsmath}
\usepackage{amsthm}
\usepackage{amssymb}
\usepackage{graphicx}
\usepackage{esint}
\usepackage{epstopdf}
\makeatletter
  \theoremstyle{definition}
  \newtheorem{example}{\protect\examplename}
  \theoremstyle{remark}
  \newtheorem{rem}{\protect\remarkname}
  \theoremstyle{definition}
  \newtheorem{defn}{\protect\definitionname}
  \theoremstyle{plain}
  \newtheorem{lem}{\protect\lemmaname}
  \theoremstyle{plain}
  \newtheorem{prop}{\protect\propositionname}
  \theoremstyle{plain}
  \newtheorem{thm}{\protect\theoremname}

\usepackage{url}
\newtheorem{assumption}{Assumption}

\makeatother

\providecommand{\definitionname}{Definition}
\providecommand{\examplename}{Example}
\providecommand{\lemmaname}{Lemma}
\providecommand{\propositionname}{Proposition}
\providecommand{\remarkname}{Remark}
\providecommand{\theoremname}{Theorem}

\begin{document}


\title{A Unified Stochastic Hybrid System Approach to Aggregate Modeling
of Responsive Loads}

\author{Lin Zhao and Wei Zhang
\thanks{This work was supported in part by the National Science Foundation
under grant ECCS-1309569 and grant CNS-1552838.}
\thanks{L. Zhao and W. Zhang (Corresponding Author) are with the Dept. of Electrical and Computer Engineering, The Ohio State University, Columbus, OH, 43210. Email:~\{zhao.833, zhang.491\}@osu.edu}
}

\maketitle
\begin{abstract}
Aggregate load modeling is of fundamental importance for systematic
analysis and design of various demand response strategies. Instead
of keeping track of the trajectories of individual loads, the aggregate
modeling problem focuses on characterizing the density evolution of the load population. Most existing models are only applicable to Thermostatically
Controlled Loads (TCL) with first-order linear dynamics. This paper develops a unified aggregate modeling approach that can be used for general TCLs as well as deferrable loads. We propose a deterministic hybrid system model to describe individual load dynamics under demand response rules, and develop a general stochastic hybrid system (SHS) model to capture the population dynamics. We also derive a set of partial differential equations (PDE) that governs the probability density evolution of the SHS. Our results cannot be obtained using the exiting SHS tools in the literature as the proposed SHS model involves both random and deterministic switchings with general switching surfaces in multi-dimensional domains. The derived PDE model includes many existing aggregate load modeling results as special cases and can be used in many other realistic modeling scenarios that have not been studied in the literature. 
\end{abstract}

\section{Introduction}

Many electric loads, such as residential HVACs (Heating, Ventilation, and Air Conditioning) and PEVs (Plug-in Electric Vehicles), can be equipped with smart local controllers that can modify their power consumptions in response to some coordination signals (e.g. price or power system frequency). Coordination of a large population of such responsive loads is an important paradigm of demand response programs and can provide various services to the power grid, such as power capping, energy arbitrage, and frequency regulation~\cite{Callaway2009,Koch2011,Mathieu2013,Zhang2013,MathieuKamgarpourLygerosEtAl2015,li2016market,TPS15_Part2}. Aggregate load modeling is concerned with characterizing the aggregate dynamics of responsive loads. The problem is of fundamental importance for systematic analysis and design of demand response strategies. It is challenging due to the various kinds of uncertainties about individual load models and the complex interplay between the load dynamics and their local response/control rules. 

One of the earliest results on aggregate load modeling was given by Malham{\'{e}} and Chong in~\cite{Malhame1985}. They considered a population of thermostatically controlled loads (TCLs), where each TCL was modeled by a two-mode stochastic hybrid system (SHS) with continuous dynamics in each mode governed by a scalar linear stochastic differential equation (SDE). Through a complicated probabilistic argument, a set of coupled Fokker-Planck equations with boundary conditions was derived, which characterizes the time evolution of the hybrid-state probability density function (p.d.f.).

The PDE (Partial Differential Equation) model derived in~\cite{Malhame1985} plays a fundamental role in the field of aggregate load modeling. Many recent studies in this area can be viewed as modifications, discretizations, approximations, or applications of the PDE model in~\cite{Malhame1985}. For example, the author in~\cite{Callaway2009}
derived the stationary solution of the coupled Fokker-Planck equations.
The solution was then utilized to perform a set point control of TCLs
for demand response. Following this work, the authors in \cite{BashashFathy2013}
derived a scalar transport equation for deterministic hybrid TCL models
using the control-volume method from continuum mechanics. The control
design was based on an ordinary differential equation (ODE) model,
which is obtained from the space-discretization of the PDE. Similar
works on deriving PDE models for TCLs were also reported in~\cite{Moura2013,Moura2014,Ghaffari2014},
where diffusion terms were used to account for parameter heterogeneity.
In particular, a linear integral output feedback control algorithm
was designed directly based on the PDE model in~\cite{Ghaffari2014}.
Moreover, Markov chain models were proposed in~\cite{Koch2011,Mathieu2013,MathieuKamgarpourLygerosEtAl2015}
to capture the evolution of the temperature distribution of a TCL
population. These models can be viewed as some approximations of the space-discretization of the underlying PDE model. 

The aforementioned works are only applicable to TCLs with linear {\em first-order} temperature dynamics. Their extensions to more complex load dynamics are not well understood. The authors in~\cite{Zhang2013} studied aggregation of second-order TCL models that involve coupled air and mass temperature dynamics. It was shown that the consideration of the second-order effect can improve the modeling performance. However, the aggregate model in~\cite{Zhang2013} was developed mainly based on heuristic arguments. In fact, when considering higher-order load dynamics, the boundary conditions of the coupled Fokker-Planck equations can be challenging to obtain, especially when there are diffusion terms in load models. In addition, many demand response applications involve deferrable loads, such as PEVs, washers, dryers, among others. These loads are dramatically different from TCLs. They are similar to computer jobs, characterized by job size and deadline. When imposing hard constraints on the deadlines, deferrable loads also exhibit dynamic behaviors.  Aggregate modeling of dynamic deferrable loads has not been adequately studied in the literature.  

This paper studies aggregate modeling of responsive loads. Different from most existing works that focus on first-order TCLs models, we develop a unified framework that can be used to obtain aggregate models for general TCLs as well as deferrable loads. In particular, we propose a general (deterministic) hybrid system model to capture individual load dynamics. The proposed hybrid system model is convenient to describe the complex multi-modal dynamics of responsive loads induced by demand response strategies. We then consider a large population of such responsive loads described by hybrid systems. To account for various kinds of uncertainties at the population level, we develop a SHS model to capture the aggregate population dynamics. Each mode of the SHS model is governed by a multi-dimensional nonlinear SDE, and mode transitions can be triggered by both deterministic and random switchings. The main technical result of this paper is the derivation of a set of coupled PDEs (later referred to as forward equations) and their boundary conditions that govern the evolution of the hybrid-state p.d.f. of the proposed SHS. The boundary conditions are particularly challenging to derive as the proposed SHS involves deterministic mode transitions characterized by a fairly general class of switching surfaces.

The proposed SHS and the derived forward equations constitute a general aggregate modeling framework for responsive loads. The main contribution of our result is two-fold. First, it provides a unified way to obtain aggregate models for a variety of responsive loads under different demand response strategies. In particular, it contains the famous result of Malham{\'{e}} and Chong~\cite{Malhame1985} as a special case. It can also be used to directly obtain the aggregate PDE models for second-order TCLs (\cite{Zhang2013,liu2016model}), which has not been formally derived in the literature. In addition, it can incorporate new scenarios such as nonlinear load dynamics, random mode switching (to account for unmodeled uncertainties on mode transitions), aggregation with deferrable loads, among others. These features can enable a larger range of applications for aggregate load modeling. 

The second contribution of this paper is on SHS. Our result, though derived in the context of aggregate load modeling, can also be used to obtain forward equations of a general SHS with both deterministic and random switchings. Although SHS has been studied extensively in the literature~\cite{Davis1993a,Bujorianu2012,HuJ2000,Blom2003,Yin2010,Bujorianu2006}, its probability density evolution cannot be explicitly characterized using the existing results, especially when there are deterministic switchings with general switching surfaces in multi-dimensional spaces. It is well known that for a standard diffusion process, the adjoint of its (strong) generator determines the form of the forward equation and the domain of the generator (indirectly) affects the boundary conditions. Unfortunately, the (strong) generator, including its domain, of a general SHS is not available in the literature. The extended generator of a generalized SHS derived in~\cite{Davis1993a,Bujorianu2006} cannot be directly used to obtain the forward equations either (See Section~\ref{sec:PDE} for technical details). Therefore, our work also contains nontrivial and important extensions of the existing theoretical results in the field of SHS.

Finally, it is worth mentioning the difference of our result with respect to two closely related works (\cite{Hespanha2005} and~\cite{Bect2010}) which also derived the forward equations for certain classes of SHS processes. The SHS model considered in~\cite{Hespanha2005} is a special case of Piecewise-Deterministic Markov Process (PDMP) with a unbounded continuous state space. It does not contain Brownian motion terms and does not involve deterministic switchings. As a result, the author does not need to derive boundary conditions, which are the key challenge of our work. Reference~\cite{Bect2010} focuses on a measure-valued formulation of the forward equation based on the Levy's identity~\cite{Walsh1972,Bass1979}. Their derivation does not provide explicit characterization of the generator boundary conditions, which, however, is the key to correctly derive the PDE boundary conditions for the forward equations.

The rest of the paper is organized as follows. Section~II provides an illustrating example for the aggregate load modeling problem. A unified hybrid system model is proposed for individual responsive
loads in Section~III. Section~IV proposes a SHS for modeling the load population considering various uncertainties. Section~V develops the  theory of the SHS
and derives the corresponding forward equation and boundary conditions.
Section~VI discusses the applications of the main theorems through
several examples. The paper is concluded in Section~VII.

\textbf{Notation}: As usual, $(\Omega,\mathcal{F},P)$ denotes the
underlying probability space of a stochastic process, where $\omega\in\Omega$
represents a sample path. $E$ denotes the expectation operation with
respect to an appropriate probability measure. For a set $U\subset\mathbb{R}^{n}$, $U^{\circ}$, $\partial U$, $\bar{U}$, and $\mathbf{1}_{U}$ represent the interior, closure, boundary, and indicator function of $U$, respectively. For a matrix $M$, $Tr(M)$, $M^{T}$, and $M_{i}$ are the trace, transpose, and $i$th row of $M$, respectively. For vectors $x,\,y\in\mathbb{R}^{n}$, we denote by $x\cdot y$ the Euclidean inner product. For a function $f:\,\mathbb{R}^{n}\mapsto\mathbb{R}$, we will use $D_{x_{i}}f$,
$f_{x_{i}}$, and $\frac{\partial f}{\partial x_{i}}$ interchangeably
for the first order partial derivatives with respect to $x_{i}$.
In addition, we use $\nabla f$ and $\nabla^{2}f$ to denote the gradient
and Hessian matrix of $f$, respectively. The divergence of a vector
field $g:\,\mathbb{R}^{n}\mapsto\mathbb{R}^{n}$ is denoted by $\nabla\cdot g$;
Similarly, the divergence of a matrix-valued function $M:\,\mathbb{R}^{n}\mapsto\mathbb{R}^{m\times n}$
is denoted by $\nabla\cdot M$, which is a vector-valued function
whose $i$th element is $\nabla\cdot M_{i}$. Let $C^{k}$ be the
class of real-valued $k$th-continuously differentiable functions
in some open subset $U\in\mathbb{R}^{n}$, and $C_{b}^{k}\subset C^{k}$
denote those functions in $C^{k}$ with bounded partial
derivatives of order up to $k$. For $f\in C^{k}$, we assume as a
convention that the value of $f$ and its $k$th order partial derivatives
on the boundary of $U$ are defined by their continuous extensions.

\section{Motivating Example \label{sec:MoEx}}
Consider a population of residential HVAC systems. Each HVAC can be described by the second-order ETP (Equivalent Thermal Parameter) model~\cite{Wilson1985}, The ETP model is a hybrid system, which has two modes, representing the ``ON/OFF'' power states of the device, and a two-dimensional continuous state vector, representing the air and mass temperatures of the house. Each mode is governed by a linear differential equation and the mode transition is triggered when the air temperature hits the boundary of the temperature dead band. Although each HVAC has simple hybrid dynamics, a large number of these loads may exhibit
rather complex aggregate dynamics under demand management
strategies. Fig. \ref{fig:EXa} shows the aggregate power response of 2000
HVACs under the so-called thermostat setback program~\cite{Kalsi2011,Zhang2013}. The parameters of the ETP model used in the simulation are generated using GridLAB-D~\cite{Zhang2013}. At the beginning of the simulation, the initial temperature setpoints of the HVACs are uniformly distributed in $[70^{\circ}\mbox{F},78^{\circ}\mbox{F}]$. During the setback event starting at time $t=3$ hour, all the setpoints are instructed to increase by $1^{\circ}\mbox{F}$, which reduces the steady state power by about $10\%$. However, after the setback control is released at time $t=6$, a large rebound is observed which may potentially damage the grid. Such phenomenon is common in demand response problems. In fact, the setpoint change described above can also be thought of as being triggered by a price change for price-responsive loads or a frequency drop for frequency-responsive loads. Therefore, developing an aggregate model that can accurately capture the collective dynamics of a population of responsive loads is of fundamental importance for demand response. This will be the main focus on this paper.  

\begin{center}
\begin{figure}
\begin{centering}
\includegraphics[clip,width=0.9\linewidth]{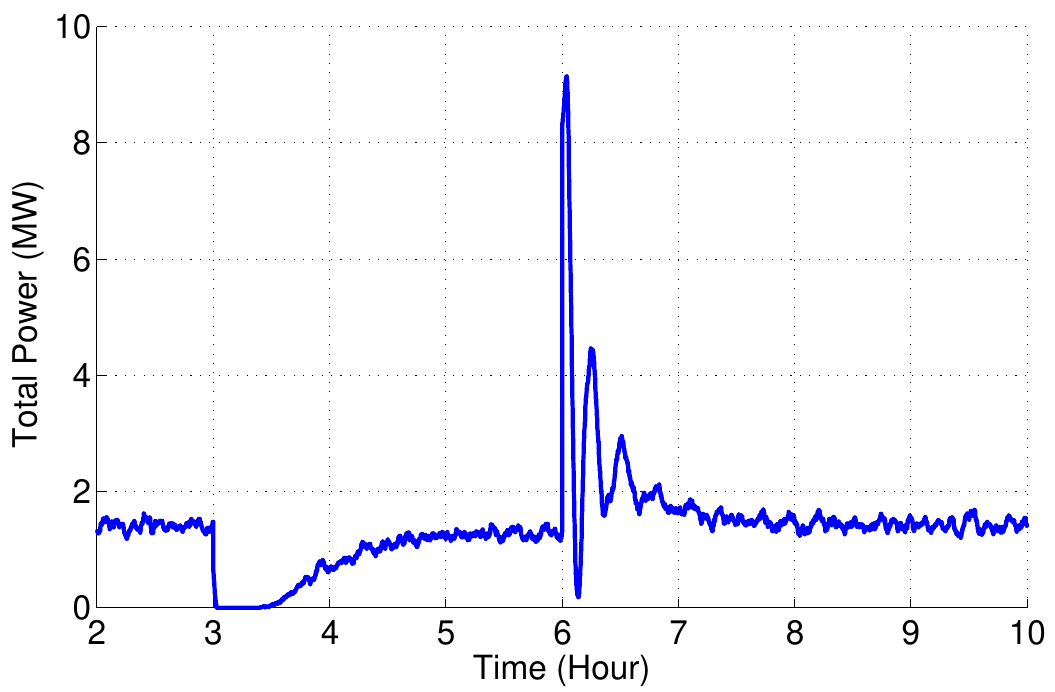}
\par\end{centering}
\caption{\label{fig:EXa}GridLAB-D simulation of thermostat setback program
of 2000 HVACs}
\end{figure}
\par\end{center}

\section{A Unified Hybrid System Model for Responsive Loads\label{sec:UHS}}
Responsive loads typically have multiple discrete operation modes.
Transitions among these modes are governed by certain switching logic
rules that depend on the evolution of the local continuous state variables (e.g. temperature of a TCL or time to completion of a PEV load)
as well as the external control signals (e.g. price or direct load control). The switching logic rules
are designed to meet end-user performances or energy consumption constraints, such as keep the temperature within a desired range or finish charging before a given deadline. To capture the interactions between the discrete and the continuous states, we propose the following hybrid system as a unified model
for responsive loads:
\begin{equation}
\begin{cases}
\dot{x}(t)=f(q(t),x(t);\theta), & \text{(continuous dynamics)}\\
q(t)=\phi(q(t^{-}),x(t^{-}),\upsilon(t^{-});\theta), & \text{(mode transition)}\\
y(t)=h(q(t);\theta), & \text{(output)}
\end{cases}\label{eq:HS_ODE}
\end{equation}
where the continuous state $x(t)\in X_{q}$ given the discrete mode $q(t)\in Q$
, and $\upsilon(t)\in\mathbb{R}$ is the \textit{external} control
input to the system which is assumed to only affect the discrete mode
transitions. We assume that $X_{q}^{\circ}$ is an open
subset of $\mathbb{R}^{n}$ with boundary $\partial X_{q}$, $Q$
is a finite subset of $\mathbb{Z}$, and $\upsilon\in\mathcal{V}$
which is the space of piecewise constant functions on $t\geq0$. We define
the hybrid state space $X\coloneqq\cup_{q}\{q\}\times X_{q}$. The
boundary of $X$ is defined as $\partial X=\cup_{q}\{q\}\times\partial X_{q}$
and the closure of $X$ is $\bar{X}:=X\cup\partial X$. For each $q\in Q$,
$f(q,\cdot;\theta)$ is a vector field of the continuous dynamics
in mode $q$. The mode transition is governed by a\emph{ transition
function} $\phi$, where $q(t^{-}):=\lim_{s\uparrow t}q(s)$ and $x(t^{-})$
and $\upsilon(t^{-})$ are defined similarly. The output function
$h(q(t);\theta)$ represents the power consumption of the load, which
typically does not depend on the continuous state. The
hybrid system model is parameterized by $\theta\in\Theta\subseteq\mathbb{R}^{n_{\theta}}$. 

The mode transition function $\phi$ is determined by the local control
logics of the responsive load. We assume that the switching logic
is characterized by switching surfaces. The external control input
$\upsilon(t)$ may directly modify the switching surfaces. We assume that the change of $\upsilon(t)$ is sufficiently slow as compared to the dynamics of the hybrid system.
To avoid further complicating the discussion, we will not
explicitly model the impact of the dynamics of $\upsilon(t)$. Nevertheless,
it will be shown through the numerical simulation. Therefore, we will fix $\upsilon(t)$ from now on. A mode transition occurs when the
continuous state $x(t)$ hits the corresponding switching surface
from within $X_{q}$. Let $\nu(q,x)$ denote the outer unit normal
vector on $\partial X_{q}$. We define the \emph{outflow switching
surface} $\mathcal{G}_{q}$ as
\begin{equation}
\mathcal{G}_{q}(\theta)=\left\{ x\in\partial X_{q}:\,f(q,x;\theta)\cdot\nu(q,x)>0\right\} .\label{eq:Gq}
\end{equation}
For the responsive loads, we assume that the continuous state remains
the same after the mode switching. In addition, we assume without
loss of generality that for each mode $q$, there are at most one
pre-jump mode, denoted by $q^{-}$, and at most one post-jump mode,
denoted by $q^{+}$. The extension to the case of multiple pre-jump
and post-jump modes is straightforward. Then we will also call $\mathcal{G}_{q}$
the \emph{inflow switching surface} to the mode $q^{+}$. Moreover,
we denote by $\mathcal{G}=\cup_{q}\{q\}\times\mathcal{G}_{q}$ the
overall outflow switching surface and $\mathcal{S}=\cup_{q}\{q^{+}\}\times\mathcal{G}_{q}$
the inflow switching surface of the hybrid system. Therefore $\phi:\,\mathcal{G}\mapsto Q$
can be simply defined as $q(t)=q^{+}$ when $x(t^{-})\in\mathcal{G}_{q}$
and $q(t^{-})=q$, otherwise $q(t)=q(t^{-})$.

The hybrid system model described above can be used to capture the
dynamics of various responsive loads under different demand response
strategies. In the following, we will give two representative examples
to illustrate its applications.
\begin{center}
\begin{figure*}
\centering{}%
\begin{minipage}[t]{0.32\textwidth}%
\begin{center}
 \includegraphics[clip,width=1\linewidth]{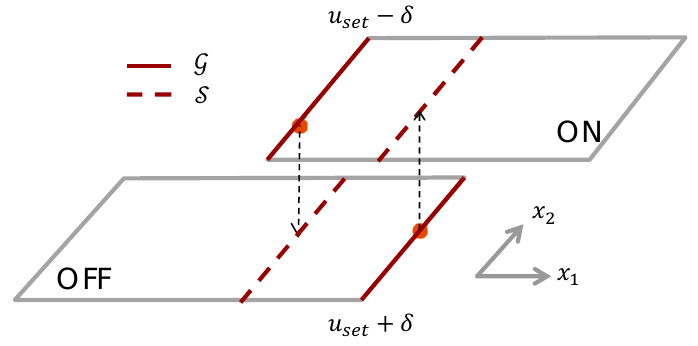} \caption{\label{fig:HVACss}Switching surfaces of the HVAC example}
\par\end{center}%
\end{minipage}\hfill{}%
\begin{minipage}[t]{0.32\textwidth}%
\begin{center}
 \includegraphics[clip,width=1\linewidth]{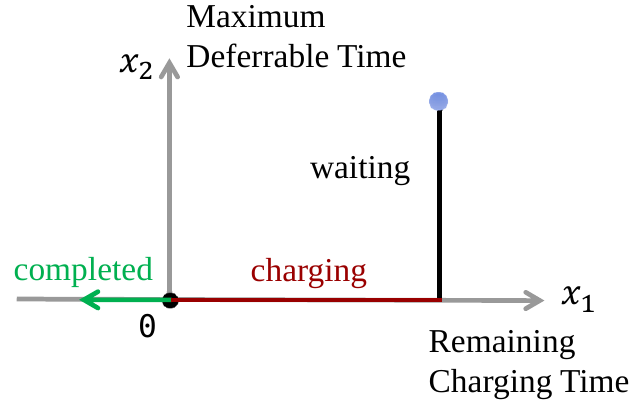} \caption{\label{fig:chargingTraj}Hybrid state trajectory of a PEV charging
load with no external control input}
\par\end{center}%
\end{minipage}\hfill{}%
\begin{minipage}[t]{0.32\textwidth}%
\begin{center}
\includegraphics[clip,width=0.8\linewidth]{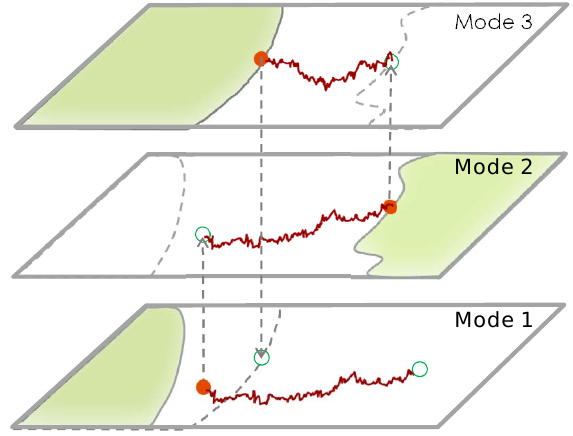}\caption{\label{fig:example}Sample trajectories of a typical SHS process}
\par\end{center}%
\end{minipage}
\end{figure*}
\par\end{center}
\begin{example}
\textbf{\label{ex: Shvac}HVACs}
\end{example}
The second-order HVAC model in the thermostat setback program described
in~\cite{Wilson1985,Zhang2013} is a special case of the proposed hybrid
sytsem~(\ref{eq:HS_ODE}). It has two discrete modes ($Q=\{0,1\}$),
representing the ``OFF'' and ``ON'' power states of the HVAC,
respectively. Each mode has a linear continuous dynamics:
\begin{equation}
\dot{x}(t)=Ax(t)+B_{q(t)},\label{eq:HVACsETP}
\end{equation}
where $x(t)=[x_{1}(t),\,x_{2}(t)]^{T}$ are the air and mass temperatures
and $q(t)\in Q$ is the discrete mode. The matrix $A$ and $B_{q}$
are determined by the ETP parameters such as  heat flux, thermal
mass of the air, thermal mass of the inner solid mass, and so on. Given a temperature
setpoint $u_{\mbox{set}}$ and a deadband size $\delta$, the continuous
state spaces are given by
\[
\begin{cases}
X_{1}=\left\{ x\in\mathbb{R}^{2}:x_{1}>u_{\mbox{set}}-\delta\right\} , & \mbox{ON mode},\\
X_{0}=\left\{ x\in\mathbb{R}^{2}:x_{1}<u_{\mbox{set}}+\delta\right\} , & \mbox{OFF mode},
\end{cases}
\]
and the corresponding switching surfaces are given by $\mathcal{G}_{1}=\partial X_{1}$
for the ON mode, and $\mathcal{G}_{0}=\partial X_{0}$ for the OFF
mode (see Fig.~\ref{fig:HVACss}). The overall system is thus parameterized
by $\theta=(A,\,B_{1},B_{2},\,u_{\mbox{set}},\,\delta)$. The system
will switch between ``ON'' and ``OFF'' upon reaching $\mathcal{G}$.
Thus the mode transition function can be written as,
\begin{equation}
q(t)=\begin{cases}
1, & x_{1}(t^{-})\geq u_{\mbox{set}}+\delta,\\
0, & x_{1}(t^{-})\leq u_{\mbox{set}}-\delta,\\
q(t^{-}), & \text{otherwise}.
\end{cases}\label{eq:localcontrol}
\end{equation}

For the thermostat setback program, the external control $\upsilon(t)$
will be the setpoint change command. It modifies the switching surfaces
and thus affects the mode transition. For price responsive loads,
$\upsilon(t)$ can also model the price signals which trigger the
setpoint change. Other TCLs under different demand response strategies
can also be modeled similarly using the proposed hybrid system framework. 
\begin{example}
\textbf{\label{ex: PEVs}Plug-in Electric Vehicle (PEV) Charging}
\end{example}
The timing dynamics of the PEV charging tasks can also be accurately
represented by the hybrid system model (\ref{eq:HS_ODE}). For example,
the timing dynamics of a PEV charging job can be described as a hybrid
system with three discrete modes $\{0,1,2\}$, representing that the
load is \textit{waiting} to be processed, being actively \textit{charging},
and has been \textit{completed}, respectively. The continuous state
is two-dimensional with $x_{1}(t)$ representing the remaining time
to finish the load if it is running, and $x_{2}(t)$ representing
the maximum time the load can be further deferred. The vector field
in each mode is given by $f(0,x)=[0,-1]^{T},$ $f(1,x)=f(2,x)=[-1,0]^{T}$.
Here, we assume that a negative value of $x_{1}(t)$ indicates that
the job has been completed before.

The continuous state spaces are given by $X_{0}=\left\{ x\in\mathbb{R}^{2}:x_{1}>0,x_{2}>0\right\} ,$
$X_{1}=\left\{ x\in\mathbb{R}^{2}:x_{1}>0\right\} ,$ and $X_{2}=\left\{ x\in\mathbb{R}^{2}:x_{1}\leq0\right\} .$
Transition from \textquotedblleft waiting\textquotedblright{} to \textquotedblleft charging\textquotedblright{}
will be forced to take place when a PEV approaches $\mathcal{G}_{0}=\{x\in\mathbb{R}^{2}:x_{1}>0,x_{2}=0\}$
from $X_{0}$ in order to meet the deadline (see Fig. \ref{fig:chargingTraj}).
A PEV finishes charging when it approaches $\mathcal{G}_{1}=\{x\in\mathbb{R}^{2}:x_{1}=0\}$
from $X_{1}$. The external control $\upsilon(t)$ may directly trigger
mode transitions at an earlier stage. For example, a frequency responsive
PEV may stop charging if the frequency deviation (which is the external
input) exceeds a certain threshold. Other deferrable loads under different
local response rules can be modeled in a similar way.

\section{Aggregate Load Modeling Using Stochastic Hybrid Systems\label{sec:Aggregation}}

Consider a large number $N$ of responsive loads, each of which is
modeled as a hybrid system of the form (\ref{eq:HS_ODE}) but may
have different parameters $\theta^{i}$, $1\leq i\le N$.
The population dynamics are described by 
\begin{equation}
\begin{cases}
\dot{x}^{i}(t)=f(q^{i}(t),x^{i}(t);\theta^{i}),\\
q^{i}(t)=\phi(q^{i}(t^{-}),x^{i}(t^{-}),\upsilon(t^{-});\theta^{i}),\\
y^{i}(t)=h(q^{i}(t);\theta^{i}),\\
y^{Aggr}(t)=\sum_{i=1}^{N}y^{i}(t),
\end{cases}\label{eq:popdyn}
\end{equation}
where $y^{Aggr}(t)$ represents the aggregate power output. The load
population is called homogeneous if all the parameters $\theta^{i}$
are the same, otherwise it is called heterogeneous.

Due to issues like privacy, limited communication resources, computational
complexities, and infrastructure costs, it is often difficult to keep
track of the hybrid state trajectory $(x^{i}(t),q^{i}(t))$ for all
the individual loads. Alternatively, it is natural to take a probabilistic
(or distribution) viewpoint for the overall population because many
information regarding individual loads is uncertain to the aggregator.
In the rest of this section, we will model the population uncertainties
and propose an SHS model for the load population.

\subsection{Population Uncertainties \label{subsec:ParaHeter}}

\subsubsection{Modeling Errors}

The continuous dynamics of responsive loads may depend on many factors
that cannot be accurately modeled. One way to account for these unmodeled
dynamics is to add a noise process to the nominal model. Therefore
we assume that the $i$th load is modeled by 
\begin{equation}
dx^{i}(t)=f(\xi^{i}(t);\theta^{i})dt+\sigma(\xi^{i}(t);\theta^{i})dW^{i}(t),\label{eq:SDE}
\end{equation}
where $\xi^{i}(t)=(q^{i}(t),x^{i}(t))$, $W^{i}(t)$ is the standard
$m$-dimensional Wiener process, and $\sigma$ is the dispersion matrix
with appropriate dimensions. In addition, we define the diffusion
matrix $\Sigma\coloneqq\sigma\sigma^{T}$.

\subsubsection{Model Parameter Uncertainties}

In contrast to a homogeneous load population, a heterogeneous one
with well-diversified load parameters will result in a natural damping
and a more stable aggregate response \cite{Callaway2009,Koch2011,BashashFathy2013}.
From the aggregator's perspective, the real model parameters $\theta^{i}$
in (\ref{eq:popdyn}) are not known precisely, but their distribution
may be acquired or estimated. Without loss of generality, we assume
$\theta\sim p_{\theta}$.

\subsubsection{Random Switching}

Users' interference can cause the spontaneous mode switchings. For
example, a user may decide to turn on/off the HVACs anytime without
abiding the local control rule. In the PEV charging example, the user
may start charging the PEVs before the deadline and drive it away
without being fully charged. This phenomenon can affect the aggregate
power output significantly. It can be modeled by the so-called random
jump mechanism~\cite{Hespanha2005,HuJ2000,Bujorianu2012}, which
is determined by a \emph{transition intensity function} $\lambda(q,x):\,Q\times\mathbb{R}^{n}\mapsto\mathbb{R}^{+}$,
where $\mathbb{R}^{+}$ denotes the non-negative real numbers. In
particular, the probability of a random jump happening in an infinitesimal
time interval $(t,t+dt]$ is given by $\lambda(q(t),x(t))dt$ .

\subsection{SHS Model for Load Population}

Although each individual load is modeled as a deterministic hybrid
system~(\ref{eq:HS_ODE}), the overall population dynamics are stochastic
due to the aforementioned uncertainties intrinsic at the aggregator
level. In this subsection, we will introduce an SHS model to describe
the aggregate population dynamics.

We assume in (\ref{eq:SDE}) that all the initial conditions $\xi^{i}(0)'s$
are i.i.d. and $\{W^{i}(t),\,1\leq i\le N\}$ are independent Wiener
processes which are also independent of $\{\xi^{i}(0),\,1\leq i\le N\}$.
Under this assumption, the loads are modeled by independent stochastic
processes. Thus if $N$ is sufficiently large, the population dynamics
can be approximated by the following SHS process, 
\begin{equation}
\begin{cases}
dx(t)=f(\xi(t),\theta(t))dt+\sigma(\xi(t),\theta(t))dW(t),\\
d\theta(t)=0,\\
q(t)=\phi(q(T_{k-1}),x(T_{k}^{-}),\upsilon(T_{k}^{-}),\theta(T_{k}^{-})),\\
\forall t\in[T_{k},T_{k+1})
\end{cases}\label{eq:HS_SDE}
\end{equation}
where $\theta(0)\sim p_{\theta}$, $\xi(0)=(q(0),x(0))\sim p_{0}$,
and we assume without loss of generality that $p_{\theta}$ and $p_{0}$
are independent. The special features of the proposed SHS model are
worth a detailed explanation. The first one is on incorporating $\theta$
as part of the continuous state with $d\theta(t)=0$. In this way,
the parameter heterogeneity can be addressed equivalently by specifying
the initial distribution $p_{\theta}$. Furthermore, the SHS incorporates
both deterministic and random mode switchings through the transition
function $\phi$, where $T_{k},\,\forall k\in\mathbb{N}$ are
random variables denoting the $k$th jump instant of the SHS process.
Specifically, $\phi:\,\bar{X}\times\mathcal{V}\times{\Theta}\mapsto Q$
can be simply defined as $q(t)=q^{+}$, $\forall t\in[T_{k},T_{k+1})$,
if $q(T_{k-1})=q$, where $T_{k}$ depends on $x,\,\upsilon,$ and
$\theta$.

The jump instants $T_{k},\,\forall k\in\mathbb{N}$ are a special
class of random variables called the stopping time. The following
definitions for $T_{k}$ extends that of the piecewise deterministic
Markov process (PDMP)~\cite{Davis1993a} to the case of the SHS process.
We first define the stopping time $t^{*}$ which triggers a deterministic
jump, and then define $T_{k}$ by incorporating the random jump. Let
$\omega^{k}:=(q^{k}(\omega),x^{k}(\omega))$ be the trajectory of
the SHS process between $[T_{k}(\omega),T_{k+1}(\omega))$, where
$x^{k}$ is the continuous component in $\omega^{k}$. Then the stopping
time $t^{*}(\omega^{k})$ which triggers a deterministic jump in mode
$q^{k}$ can be defined as
\[
t^{*}(\omega^{k})=\inf\,\left\{ t>0;\,x^{k}(t,\omega)\in\mathcal{G}_{q^{k}}\right\} .
\]

Now we incorporate the random jump. Let $S^{k}$ be the dwell time
of the process in mode $q^{k}$, which is defined by 
\begin{equation}
S^{k}(\omega^{k+1};\omega^{k})=\inf\{t:\,\beta(t;\omega^{k})\leq u(\omega^{k+1})\},\label{eq:stopping}
\end{equation}
where $\beta$ is the survival function of $S^{k}$ such that $\beta(t)=P(S^{k}>t)$,
and $u$ is a uniformly distributed random variable between $[0,1]$.
The survival function is given by~\cite{Davis1993a},
\begin{equation}
\beta(t;\omega^{k})=I_{\{t<t^{*}(\omega^{k})\}}\exp\left(-\int_{0}^{t}\lambda(q^{k},x^{k}(s,\omega))ds\right),\label{eq:Survival}
\end{equation}
where $\lambda$ is the transition intensity function. Thus the mode
transition of the SHS process occurs at the following stopping times
\begin{equation}
T_{k+1}=T_{k}+S^{k},\,T_{0}=0.\label{eq:StoppingTk}
\end{equation}
Without loss of generality, we assume that $\lambda(q,x)$ is piecewise
continuous, $\forall q\in Q$. This is to ensure that $S^{k}>0$ almost
surely (a.s.). Furthermore, it can be seen that the longer the process
has been staying in a mode, the more likely the mode will change. 

A sample path of (\ref{eq:HS_SDE}) can be viewed as the trajectory
of a randomly selected load in the population. Fig.~\ref{fig:example}
illustrates the sample path of a typical SHS process with three mode.
The boundary curves of the shaded area represent the outflow switching
surfaces, and the dashed curves represent the inflow switching surfaces.
We see that a random jump happens in mode 1 before the load reaches
the outflow switching surface, and in the other two modes, deterministic
jump happens upon reaching the outflow switching surfaces. Suppose
$p(q,x,t)$ is the p.d.f. of the hybrid state of the SHS process,
and $Q_{0}\subset Q$ are the modes under which the loads are
consuming energy. Then the aggregate power response can be calculated
by 
\[
y^{Aggr}(t)=N\sum_{q\in Q_{0}}\int_{\Theta}\int_{X_{q}}h(q,\theta)p(q,x,t)p_{\theta}dxd\theta.
\]

\section{PDE Characterization of Aggregate Dynamics}
\label{sec:PDE}
As discussed in the previous section, the load population can be modeled
as an SHS of the form~(\ref{eq:HS_SDE}). For systematic analysis,
it is often desired or even necessary to characterize the evolution
of the hybrid-state probability density function of the SHS, which
is the main focus of this section. In fact, most of the existing aggregate
load models in the literature are essentially characterizing the density
evolutions of some simplified versions of our general SHS model~(\ref{eq:HS_SDE}).

Our main results will be derived for the case of homogeneous parameters
first. The extension to the case of heterogeneous parameters will
be discussed at the end of this section. We therefore drop the dependence
of $\theta$ for now and focus on the following SHS,
\begin{multline}
\begin{cases}
dx(t)=f(q(t),x(t))dt+\sigma(q(t),x(t))dW(t),\\
q(t)=\phi(q(T_{k-1}),x(T_{k}^{-}),\upsilon(T_{k}^{-})),\\
\forall t\in[T_{k},T_{k+1})
\end{cases}\label{eq:SDE3}
\end{multline}
where note that the definition of $T_{k}$ depends on the transition
intensity function $\lambda$ and the outflow switching surface $\mathcal{G}$. 

For a standard diffusion process, its density evolution is characterized
by the Fokker-Planck equation, which can be derived using the infinitesimal
generator and the associated Dynkin's formula~\cite{Schuss2010}.
A similar idea can be used to derive a set of coupled PDEs characterizing
the density evolution for a switching diffusion process~\cite{Malhame1985}.
However, characterizing the density evolution for the general SHS
model~(\ref{eq:SDE3}) is much more challenging. The main challenge
lies in the deterministic switchings that are forced to occur when
the continuous state hits certain switching surfaces. These switchings,
along with the possible random switchings and diffusion noises, significantly
complicate the boundary conditions of the resulting density PDEs.
In the rest of the section, we will first review some important concepts
and the key steps in deriving forward equations of a general Markov
process. Then we will generalize the existing results and derive the
density evolution PDEs for the homogeneous SHS~(\ref{eq:SDE3}).
The last subsection will introduce a method to deal with the general
heterogeneous case. 

\subsection{Preliminaries on Markov Processes}

We first review some classical concepts and results that are useful
in deriving the Kolmogorov forward equations of a general Markov process.
The discussion in this subsection will not only set up stages for
our main results, but also allow us to see the subtle technical challenges
in deriving the forward equations for the general SHS model.

Let $\xi_{t}$, $t\geq0$ be a (time homogeneous) Markov process defined
on $X$, whose transition probability measure is denoted by $\varrho:\,\mathbb{R}^{+}\times X\times\mathcal{B}(X)\mapsto[0,1]$.
Let $P_{\xi_{0}}$ be the Wiener probability measure of $\xi_{t}$
such that $P_{\xi_{0}}(\xi_{t}\in A)=\varrho(t,\xi_{0},A)$, $\forall A\in\mathcal{B}(X)$.
Given $\mu$ the initial distribution of $\xi_{0}$, the abstract
probability measure $P$ can be related to $P_{\xi_{0}}$ by $P=\int_{X}P_{\xi_{0}}d\mu$ . If $\mu$
and $P$ admits a probability density function $p_{0}$ and $p$
respectively, then we have the relation $d\mu=p_{0}(\xi)d\xi$
and $dP=p(\xi)d\xi$.

For a real-valued bounded Borel measurable function $\psi$ on $X$,
we can define the semi-group $Z_{t}\psi:=E_{\xi_{0}}\psi(\xi_{t})$,
$t\geq0$, where the expectation $E_{\xi_{0}}$ is with respect to
$P_{\xi_{0}}$. The (infinitesimal or strong) generator of the process
$\xi_{t}$ is defined as follows:
\begin{defn}
(Strong Generator) Let $\mathcal{D}(L)$ denote the set of bounded
Borel measurable functions $\psi:$$X\mapsto\mathbb{R}$ with the
property that $\forall\psi\in\mathcal{D}(L),$ the limit 
\begin{equation}
L\psi\coloneqq\lim_{t\downarrow0}\frac{Z_{t}\psi-\psi}{t},\label{eq:DefStrgGen}
\end{equation}
exists in the supremum norm $\left\Vert \cdot\right\Vert \coloneqq\sup_{\xi\in X}$$\left|\cdot\right|$.
Then we call $(L,\mathcal{D}(L))$ the strong generator of the process
$\xi_{t}$.
\end{defn}
Furthermore, from the Markov process theory~\cite{Davis1993a}, we
know:
\begin{lem}
\label{lem:stronggen}Each $\psi\in\mathcal{D}(L)$ is associated
with a martingale defined as,
\begin{equation}
C_{t}^{\psi}=\psi(\xi_{t})-\psi(\xi_{0})-\int_{0}^{t}L\psi(\xi_{s})ds,\label{eq:martingale}
\end{equation}
which satisfies the Dynkin's formula,
\begin{equation}
E_{\xi_{0}}\psi(\xi_{t})=\psi(\xi_{0})+E_{\xi_{0}}\left[\int_{0}^{t}L\psi(\xi_{s})ds\right].\label{eq:Dynkin}
\end{equation}
\end{lem}
The Dynkin's formula defines the time evolution of the expectation
of a function $\psi$ of the stochastic processes. It plays a major
role in deriving the PDE model. Let $C_{c}^{\infty}(X)$ denote the
space of real-valued smooth functions on $X$ with compact support, $\left\langle ,\right\rangle $ denote the inner product of $L_{2}(X)$
of square integrable functions on $X$. For a (linear) differential
operator $V$, we will say $V^{*}$ is the $L_{2}$ \emph{formal}
\emph{adjoint }of $L$ if
\begin{equation}
\bigl\langle V\psi_{1},\psi_{2}\bigr\rangle=\bigl\langle\psi_{1},V^{*}\psi_{2}\bigr\rangle,\label{eq:adjoint}
\end{equation}
for all smooth functions $\psi_{1}$, $\psi_{2}\in C_{c}^{\infty}(X)$.
Note that $V^{*}$ can be calculated by shifting the differential
operator from $\psi_{1}$ to $\psi_{2}$ using integration by
parts or divergence theorem (see Lemma~\ref{lem:divergence}
in the Appendix~\ref{subsec:ProofThm1}). 
\begin{lem}
\label{lem:CalculateAdjoint}Let $(L,\mathcal{D}(L))$ be the strong
generator of the Markov process $\xi_{t}$ on $X$ and $p(\xi,t)$
be its probability density function. Suppose $L$ is a differential
operator and $X$ has a piecewise smooth boundary. Then, in the sense
of weak derivatives, $p(\xi,t)$ satisfies the PDE 
\[
\frac{\partial p}{\partial t}=L^{*}p,\,t\geq0,
\]
with boundary conditions uniquely determined by $\mathcal{D}(L)$.
\end{lem}
\begin{IEEEproof}
For smooth functions $\mathcal{\psi}\in\mathcal{D}(L)$, using
the divergence theorem, we can write formally 
\[
\bigl\langle\frac{\partial p}{\partial t}-L^{*}p,\psi\bigr\rangle=\bigl\langle\frac{\partial p}{\partial t},\psi\bigr\rangle-\bigl\langle p,L\psi\bigr\rangle+\Gamma(\psi,p),
\]
where $\Gamma$ is the surface integral on $\partial X$ induced by
the differential operator $L$. Then the PDE holds in the sense of
weak derivatives~\cite{Evans2010}, if 
\begin{equation}
\begin{cases}
\bigl\langle\frac{\partial p}{\partial t},\psi\bigr\rangle=\bigl\langle p,L\psi\bigr\rangle,\\
\Gamma(\psi,p)=0,
\end{cases}\label{eq:WeakSol}
\end{equation}
for all the smooth functions $\mathcal{\psi}\in\mathcal{D}(L)$. Let $\Lambda(L)\supset\mathcal{D}(L)$
denote the set of the boundary conditions satisfied by $\psi\in\mathcal{D}(L)$, then
the boundary conditions $\Lambda(L^{*})$ is defined as the minimal
set of (homogeneous) conditions such that $\Gamma(\psi,p)=0$ for
all $\psi\in\Lambda(L)$ and $p\in\Lambda(L^{*})$ (see~\cite[page 103]{DuChateauZachmann1986}).
Note that $\Lambda(L^{*})$ is uniquely determined by $\Lambda(L)$
by this definition.

Furthermore, we know from Lemma~\ref{lem:stronggen} that the Dynkin's
formula holds for all $\mathcal{\psi}\in\mathcal{D}(L)$. Taking the
time derivative of both sides of~(\ref{eq:Dynkin}), we have $\bigl\langle\frac{\partial p}{\partial t},\psi\bigr\rangle-\bigl\langle p,L\psi\bigr\rangle=0$.
Thus it completes the proof.
\end{IEEEproof}
We see from Lemma~\ref{lem:CalculateAdjoint} that the PDE is determined
by the formal adjoint operator of $L$, and the associated boundary
conditions are uniquely determined by the boundary conditions in $\mathcal{D}(L)$.
Lemma~\ref{lem:CalculateAdjoint} has been used to obtain the PDEs
for diffusion processes~\cite{Schuss2010} on subsets of $\mathbb{R}^{n}$
with different kinds of boundaries. Similar idea has also been used
to derive the PDE for the jump-diffusion processes~\cite{Hanson2004}.

For the SHS process defined on a hybrid state space $X$,
characterizing the strong generator $(L,\mathcal{D}(L))$ can be very
difficult. However, for the purpose of deriving the PDE, we can instead
characterize an extended operator of the strong generator. Before
proceeding, we will introduce the following assumptions for the later discussion.

\subsection{Standing Assumptions}

In this section, we will introduce some regularity conditions. The
first two assumptions are imposed to ensure that the SHS processes
behave nicely, while Assumptions~\ref{Ass:mbd} and \ref{Ass:Inside}
are introduced to facilitate the derivation of the PDE. 

Denote by $\xi_{t}=\xi(t)=(q(t),\,x(t))$ the hybrid state of the
SHS. Let $f_{i}$ be the $i$th entry of $f$, and $\sigma_{ij}$
be the $(i,j)$-entry of $\sigma$. For a function $\psi$ defined
on the hybrid state space $X$, we denote $\psi^{q}=\psi\left|_{X_{q}}\right.$
the restriction of $\psi$ to $X_{q}$. 

\begin{assumption}\label{Ass:regularity}

For the SDE in (\ref{eq:SDE3}), we assume that $f_{i}^{q}\in C_{b}^{1}$,
$\sigma_{ij}^{q}\in C_{b}^{1}\cap C^{2}$, $\forall q\in Q$. In addition,
the initial state $\xi_{0}$ is independent of $\{W(t),\,t\geq0\}$
and satisfies $E(\left\Vert \xi_{0}\right\Vert ^{2})<\infty$.

\end{assumption}
\begin{rem}
\label{rem:Remark1}This is a standard assumption to guarantee that
the initial value problem of (\ref{eq:SDE3}) has a unique continuous
solution (i.e., the It{\^{o}} diffusion), and the solution satisfies
$E(\int_{0}^{t}\left\Vert x(s)\right\Vert ^{2}ds)<\infty$, $\forall t>0$
\cite[Theorem 5.2.1]{Oksendal2003}. The smoothness of $f$ and $\sigma$
is imposed for obtaining a PDE model in the classical sense.
\end{rem}
\begin{assumption}\label{Ass:NoZeno}

There is no Zeno execution for the hybrid system of (\ref{eq:SDE3}).

\end{assumption}
\begin{rem}
\label{rem:Nt}Let $N_{t}(\omega)\coloneqq\sup\,\{k\in\mathbb{N}:T_{k}(\omega)\leq t\}$,
which is the number of jumps happened before time $t$. Then Assumption~\ref{Ass:NoZeno}
implies $E_{\xi_{0}}(N_{t})<\infty$, $\forall\xi_{0}\in X$. This
is a common assumption for the SHS.
\end{rem}
Regarding the structure of the hybrid state space $X$ of the SHS
process, we have,

\begin{assumption}\label{Ass:mbd}

$\forall q\in Q$, $\bar{X}_{q}$ is a connected and oriented $C^{2}$
manifold with corners.

\end{assumption}

The above assumption requires that the state space is locally $C^{2}$-diffeomorphic
to $[0,\infty)^{n}$. It unifies the description of the state spaces
of both the TCLs (e.g. HVACs) and defferable loads (e.g. PEV) and
facilitates our proof later. More details on manifold with corners
can be found in~\cite{Lee2013}.

Recall that $\mathcal{G}_{q}$ is the outflow switching surface in
mode $q$ and $\mathcal{G}\coloneqq\cup_{q}\{q\}\times\mathcal{G}_{q}.$
For the SHS process, we can define $\mathcal{G}_{q}$ explicitly as
\begin{multline}
\mathcal{G}_{q}:=\Bigl\{ x\in\partial X_{q}:\,f(q,x)\cdot\nu(q,x)>0\\
\text{or }\nu^{T}(q,x)\Sigma(q,x)\neq\boldsymbol{0}\Bigr\},\label{eq:GqSHS}
\end{multline}
where $\nu(q,x)$ is the outer unit normal vector on $\partial X$.
This extends the definition in~(\ref{eq:Gq}) for the deterministic
system. It is possible that the outflow switching surface $\mathcal{G}_{q}$
may not contain all of $\partial X_{q}$, i.e., not the entire boundary
can be reached from some interior point of $X_{q}$. For example,
in the deterministic case (i.e., $\sigma\equiv0$), $\mathcal{G}_{q}$
is the forward reachable boundary defined in~(\ref{eq:Gq}). Moreover,
we assume that:

\begin{assumption}\label{Ass:Inside}

The outflow switching surface $\mathcal{G}_{q}$ is an open subset
of $\partial X_{q}$.

\end{assumption}
\begin{rem}
Assumptions \ref{Ass:mbd} and \ref{Ass:Inside} imply that $\mathcal{G}_{q}$
is also oriented~\cite[Chapter 15]{Lee2013}. This is to guarantee
that the integrals over manifolds are defined consistently. In particular,
this is a standing assumption of the divergence theorem that will
be used in the proof of our Theorem~\ref{thm:MainPDE}.
\end{rem}
The following defines a metric on the hybrid state space $X$~\cite[page 58]{Davis1993a}.
A distance function between the hybrid states can be defined by $\rho(\xi,\xi^{'})=1$
if $q\neq q'$ and $\rho(\xi,\xi^{'})=\chi(x-x')$ if $q=q'$, where
$\chi(x)=\frac{2}{\pi}\tan^{-1}(\left\Vert x\right\Vert )$, $\forall x\in\mathbb{R}^{n}$.
Now $X$ can be endowed with the Borel $\sigma-$algebra $\mathcal{B}(X)$
generated by its metric topology, where $\mathcal{B}(X)=\sigma\left\{ \cup_{q}\left\{ q\right\} \times\mathcal{B}_{q}\right\} $
and $\mathcal{B}_{q}$ is the $\sigma-$algebra on $X_{q}$.

For clarity, we introduce the following assumptions
on the switching surfaces and define the partitions of the hybrid
state space. Since the SHS process is defined on $X$, we assume that
the post-jump position $(\phi(q,x),x)\in X$, for a jump from $(q,x)\in\bar{X}$.
Recall that $\mathcal{S}\coloneqq\cup_{q}\{q^{+}\}\times\mathcal{G}_{q}$
is the union of all the inflow switching surfaces. Therefore, we have
$\mathcal{S}\subset X$. In addition, it is assumed that $\mathcal{S}\cap\mathcal{G}=\emptyset$.
However, note that $\mathcal{S}\cap\partial X$ may not be empty.
Moreover, define the surface $\mathcal{E}\coloneqq\cup_{q}\{q^{+}\}\times\partial X_{q}$.
Clearly, by Assumption~\ref{Ass:Inside}, $\mathcal{S}$ is an open
subset of $\mathcal{E}$. We assume that $\mathcal{E}$ forms a partition
of $\bar{X}$, that is, $\bar{X}=\cup_{q\in Q,i\in\mathbb{N}_{q}}\{q\}\times\bar{X}_{q}^{i}$,
such that the each\emph{ partition} $\{q\}\times X_{q}^{i}$ satisfies
$\{q\}\times(\bar{X}_{q}^{i}\cap\bar{X}_{q}^{j})\subset\mathcal{E}$,
$\forall q\in Q$, $i,j\in\mathbb{N}_{q}$, $i\neq j$, where $\mathbb{N}_{q}$
is the index set of the partitions of $\{q\}\times X_{q}$.

\subsection{Dynkin's Formula for the SHS Processes}

As discussed before, the PDE is completely determined by the strong
generator $(L,\mathcal{D}(L))$. However, generally it can be very difficult
to directly characterize $\mathcal{D}(L)$. Note that in order to
apply the idea of Lemma~\ref{lem:CalculateAdjoint}, we only need
the Dynkin's formula hold, and it might be easier to characterize
the generator boundary condition by characterizing a super set of
$\mathcal{D}(L)$. To this end, we introduce the notion of the extended
generator (c.f.~\cite[Definition 14.15]{Davis1993a}).
\begin{defn}
\label{def:ExGen}(Extended Generator) Let $\mathcal{D}(\hat{L})$
denote the set of Borel measurable functions $\psi:X\mapsto\mathbb{R}$
with the following property: there exists a measurable function $h:\,X\mapsto\mathbb{R}$,
such that the function $t\mapsto h(\xi_{t})$ is integrable $P_{\xi_{0}}$
a.s. for each $\xi_{0}\in X$ and the process 
\[
C_{t}^{\psi}=\psi(\xi_{t})-\psi(\xi_{0})-\int_{0}^{t}h(\xi_{s})ds,
\]
is a local martingale. Then we write $h=\hat{L}\psi$ and call $(\hat{L},\mathcal{D}(\hat{L}))$
the extended generator of the process $\xi_{t}$.
\end{defn}
Since a martingale is also a local martingale, we see that $\mathcal{D}(L)\subset\mathcal{D}(\hat{L})$,
and $\hat{L}\psi=L\psi$ for $\psi\in\mathcal{D}(L)$, and therefore
the name ``extended generator''. More importantly, it can be easily
verified that set $\mathcal{D}_{m}:=\{\psi\in\mathcal{D}(\hat{L}):\,C_{t}^{\psi}\,\text{is a martingale}\}$
is precisely the largest class of functions for which the Dynkin's
formula~(\ref{eq:Dynkin}) holds. Clearly, we have $\mathcal{D}(L)\subset\mathcal{D}_{m}\subset\mathcal{D}(\hat{L})$.
Since our purpose is to derive the PDE
using the Dynkin's formula, it is our major focus to characterize
$\mathcal{D}_{m}$ now.

Before proceeding, we introduce the following notations~\cite{Davis1993a}.
For a measurable function $g:\,X\times\mathbb{R}_{+}\mapsto\mathbb{R}$,
we say $g(\xi,s)\in L_{1}(\tau)$ if $E(\int\left|g\right|d\tau)<\infty$,
where $\tau$ is a stochastic process that can be identified by a
random measure and thus defines a Stieljes integral in the expression.
Typically, we will denote $\tau$ as the counting process defined
by
\begin{equation}
\tau(t,A)\coloneqq\sum\mathbf{1}_{\{T_{k}\leq t\}}\mathbf{1}_{\{\xi_{T_{k}}\in A\}},\label{eq:countingP}
\end{equation}
which counts the number of jumps of the GSHS process $\xi_{t}$. Then
we say $g\in L_{1}^{\mbox{loc}}(\tau)$ if there is a sequence of
stopping times $\gamma_{n}$ with $\gamma_{n}\uparrow\infty$ a.s.
such that $g\mathbf{1}_{\{s\leq\gamma_{n}\}}\in L_{1}(\tau)$ for
$\forall n\in\mathbb{N}_{+}.$ For $\psi:\,X\mapsto\mathbb{R}$, let
$B\psi:\,X\times\mathbb{R}_{+}\times\Omega\mapsto\mathbb{R}$ be defined
as $B\psi(\xi,s,\omega)\coloneqq\psi(\xi)-\psi(\xi_{s^{-}}(\omega)).$
The function $B\psi$ is introduced to evaluate the increments of
$\psi$ due to the jumps. Finally, we assume as a convention that
for a continuous function on $X$, its value on the boundary $\mathcal{G}$
is defined by its continuous extension.

We first cite a characterization result of the extended generator
of a class of general SHS (GSHS) processes proposed in~\cite{Bujorianu2012}.
\begin{lem}
\label{lem:BM}(GSHS extended generator~\cite[Theorem 4.11]{Bujorianu2012})
The extended generator $(\hat{L},\mathcal{D}(\hat{L}))$ of a GSHS
process $\xi_{t}$ satisfies $\forall\psi\in\mathcal{D}(\hat{L}),$
\begin{multline}
(\hat{L}\psi)(\xi)=\nabla\psi(\xi)\cdot f(\xi)+\frac{1}{2}Tr(\Sigma(\xi)\nabla^{2}\psi(\xi))\\
+\lambda(\xi)\int_{X}\left(\psi(\zeta)-\psi(\xi)\right)\mathcal{R}(\xi,d\zeta),\label{eq:GSHSGen}
\end{multline}
with $\mathcal{D}(\hat{L})$ including at least those functions $\psi:\,X\mapsto\mathbb{R}$
such that:

1. $\psi^{q}\in C^{2},\,\forall q\in Q$.

2. (boundary condition) $\psi(\xi)=\int_{X}\psi(\zeta)\mathcal{R}(\xi,d\zeta),\ \xi\in\mathcal{G}$.

3. $B\psi\in L_{1}^{\mbox{loc}}(\tau)$.

\noindent where $\mathcal{R}:\bar{X}\times\mathcal{B}(X)\mapsto[0,1]$
is a transition measure on the post-jump positions given the pre-jump
position.
\end{lem}
\begin{rem}
An exact characterization of the extended generator has been obtained
for the PDMP~\cite[Theorem (26.14)]{Davis1993a}. It hinges on the
local martingale representation theorem for\emph{ the general jump
process}. However, such a theorem is unavailable for the GSHS process,
and only the sufficiency of these conditions can be claimed. In Lemma~\ref{lem:BM},
the key information of the boundary condition is obtained through
characterizing the extended generator. It is worth mentioning that
this is the same boundary condition satisfied by the extended generator
of the PDMP.

Next, we will characterize a subset of $\mathcal{D}(\hat{L})$ on
which $C_{t}^{\psi}$ is actually a martingale (rather than a local
martingale) with respect to the natural filtration $\mathcal{F}_{t}$
generated by the GSHS process.
\end{rem}
\begin{lem}
(Martingale characterization)\label{lem:L1Mart}Let $(\hat{L},\mathcal{D}(\hat{L}))$
be the extended generator of a GSHS process $\xi_{t}$ and $\psi\in\mathcal{D}(\hat{L})$.
If $\psi^{q}$ is bounded $\psi^{q}\in C_{b}^{1}$ for all $q\in Q$,
then $C_{t}^{\psi}$ is an $\mathcal{F}_{t}$-martingale.
\end{lem}
\begin{IEEEproof}
See Appendix~\ref{subsec:Proof1}.
\end{IEEEproof}
Based on Lemmas~\ref{lem:BM} and~\ref{lem:L1Mart}, we are now
in the position to establish the Dynkin's formula for the proposed
SHS process in~(\ref{eq:SDE3}). Note that the post-jump position
of the proposed SHS process is specified by the mode transition function
$\phi$ deterministically, that is, there is a one-to-one correspondence
between the pre-jump position and post-jump position. This will give
us a local boundary condition for the extended generator.
\begin{prop}
\label{prop:Lf}(SHS extended generator) Under the Assumptions~\ref{Ass:regularity}
and \ref{Ass:NoZeno}, the extended generator $(\hat{L},\mathcal{D}(\hat{L}))$
of the SHS process~(\ref{eq:SDE3}) satisfies $\forall\psi\in\mathcal{D}(\hat{L})$,
\begin{multline}
(\hat{L}\psi)(q,x)=\nabla\psi(q,x)\cdot f(q,x)+\frac{1}{2}Tr(\Sigma(q,x)\nabla^{2}\psi(q,x))\\
+\lambda(q,x)\left(\psi(q^{+},x)-\psi(q,x)\right),\label{eq:Generator}
\end{multline}
with $\mathcal{D}(\hat{L})$ including at least those functions $\psi:\,X\mapsto\mathbb{R}$
such that: (1) $\psi^{q}\in C^{2},\,\forall q\in Q$, (2) (boundary
condition) $\psi(q^{+},x)=\psi(q,x),\,\forall(q,x)\in\mathcal{G}$,
and (3) $B\psi\in L_{1}^{\mbox{loc}}(\tau)$.
\end{prop}
\begin{IEEEproof}
See Appendix~\ref{subsec:Proofcor1}.
\end{IEEEproof}
For all $\psi\in\mathcal{D}(\hat{L})$ characterized in Proposition~\ref{prop:Lf},
$C_{t}^{\psi}$ is a local martingale on the SHS process~(\ref{eq:SDE3}).
We define,
\begin{multline*}
\mathcal{D}=\Bigl\{\psi:\,X\mapsto\mathbb{R}\Bigl|\psi\text{ is bounded},\,\psi^{q}\in C_{b}^{1}\cap C^{2},\\
\text{and satisfies }\psi(q^{+},x)=\psi(q,x),\,\forall(q,x)\in\mathcal{G}\Bigr\}.
\end{multline*}
Then combining Lemma~\ref{lem:L1Mart} and Proposition~\ref{prop:Lf},
we can obtain:
\begin{prop}
\label{prop:DynkinCt}(SHS Dynkin's formula) For all $\psi\in\mathcal{D}$,
$C_{t}^{\psi}$ is a martingale and the Dynkin's formula (\ref{eq:Dynkin})
holds.
\end{prop}
\begin{IEEEproof}
See Appendix~\ref{subsec:ProofCorDynkinCt}.
\end{IEEEproof}
Note that the characterized set $\mathcal{D}$ is a subset of $\mathcal{D}_{m}$.
However, it contains enough functions for deriving the PDE model.

\subsection{PDE Model}

In this subsection, we derive the PDE model based on the Dynkin's
formula (\ref{eq:Dynkin}) proved in Proposition~\ref{prop:DynkinCt}. 

In view of the partition of the state space $X$, we will start by
assuming that the $p(q,x,t)$ is piecewise $C^{2,1}$ in the interior
of $X_{q}\times\mathbb{R}^{+}$, $\forall q\in Q$, where $C^{2,1}$
means that $p(q,x,t)$ is $C^{2}$ with respect to $x$ and $C^{1}$
with respect to $t$. Of course, certain smoothness conditions on
the initial distribution and coefficients are required to make this
happen, otherwise $p$ is only a weak solution in the sense of~(\ref{eq:WeakSol})
(i.e., in the sense of weak derivatives, see~\cite{Evans2010}).
The detailed discussion will be lengthy and out of the scope of this
paper. Hence, we will restrict ourselves to the smooth assumption.

Let us denote $p(q,x,t)=p_{i}(q,x,t)$ in $X_{q}^{i}$, $\forall q\in Q$,
$i\in\mathbb{N}_{q}$.
\begin{thm}
\label{thm:MainPDE}(PDE characterization) Under the Assumptions~\ref{Ass:regularity}-\ref{Ass:Inside},
for all $q\in Q$, $i\in\mathbb{N}_{q}$, $x\in X_{q}^{i}$, the hybrid-state
probability density function $p_{i}(q,x,t)$ satisfies the following
PDE:
\begin{equation}
T[p_{i}]=0,\label{eq:FKPDE}
\end{equation}
where
\begin{align*}
T[p]= & \frac{\partial p(q,x,t)}{\partial t}+\mbox{\emph{\ensuremath{\nabla}}}\cdot\gamma(q,x,t)\\
 & +\lambda(q,x)p(q,x,t)-\lambda(q^{-},x)p(q^{-},x,t),
\end{align*}
and $\gamma$ is known as the probability flux given as
\[
\gamma(q,x,t)=f(q,x)p(q,x,t)-\frac{1}{2}\emph{\ensuremath{\nabla}}\cdot(p(q,x,t)\Sigma(q,x)).
\]
\end{thm}
\begin{IEEEproof}
See Appendix~\ref{subsec:ProofThm1}.
\end{IEEEproof}
We see that the above PDEs are coupled in the state spaces $X_{q}^{i}$
due to the source terms $\lambda p$ caused by the random jumps. Noting
that without the random jump, i.e. $\lambda\equiv0$, equation (\ref{eq:FKPDE})
reduces to the well-known Fokker-Planck equations. Moreover, they
are also coupled on the boundaries $\mathcal{G}$ and $\mathcal{S}$
due to the deterministic jumps. Let $\nu(q,x)$ denote the outer unit
normal vector on $\mathcal{G}_{q}$, and define $\overrightarrow{p}(q^{+},x,t):=\lim_{\varepsilon\uparrow0}p(q^{+},x+\varepsilon\nu,t)$
and $\overrightarrow{\gamma}(q^{+},x,t):=\lim_{\varepsilon\uparrow0}\gamma(q^{+},x+\varepsilon\nu,t)$.
These are respectively the continuous extensions of the probability
flux $p$ and density $\gamma$ on $\mathcal{S}$ from the side of
$\mathcal{S}$ that has the same outer normal direction as $\mathcal{G}$.
Similarly, define $\overleftarrow{\gamma}(q^{+},x,t):=\lim_{\varepsilon\downarrow0}\gamma(q^{+},x+\varepsilon\nu,t)$
and $\overleftarrow{p}(q^{+},x,t):=\lim_{\varepsilon\downarrow0}p(q^{+},x+\varepsilon\nu,t)$
the quantities continuously extended from the opposite side of $\mathcal{S}$.
We have:
\begin{thm}
\label{thm:MainBC}(PDE boundary conditions) The PDE model in (\ref{eq:FKPDE})
satisfies the following boundary conditions, for all $q\in Q,$ \textup{$x\in\mathcal{G}_{q},$}
\begin{equation}
p(q,x,t)\nu^{T}(q,x)\Sigma(q,x)=\mathbf{0},\label{eq:Absorbing}
\end{equation}
\begin{equation}
\left(\overrightarrow{p}(q^{+},x,t)-\overleftarrow{p}(q^{+},x,t)\right)\nu^{T}(q,x)\Sigma(q^{+},x)=\mathbf{0},\label{eq:Continuity}
\end{equation}
\begin{equation}
\left[\overrightarrow{\gamma}(q^{+},x,t)-\overleftarrow{\gamma}(q^{+},x,t)+\gamma(q,x,t)\right]\cdot\nu(q,x)=0.\label{eq:Conservation}
\end{equation}
\end{thm}
\begin{IEEEproof}
See Appendix~\ref{subsec:ProofThm2}.
\end{IEEEproof}
A closer look at~(\ref{eq:Absorbing})-(\ref{eq:Conservation})
will give us insight into the patterns of the boundary conditions. Physically,
$\nu^{T}(q,x)\Sigma(q,x)$ represents the projection of the multi-dimensional
Brownian motion along the normal direction of the outflow switching
surface. In particular, if $\nu^{T}(q,x)\Sigma(q,x)\neq\mathbf{0}$,
then
\begin{enumerate}
\item Condition~(\ref{eq:Absorbing}) implies that the p.d.f. at any point
of the outflow switching surfaces must be $0$; 
\item Condition~(\ref{eq:Continuity}) implies that the p.d.f. is continuous
across the inflow switching surfaces;
\item Condition~(\ref{eq:Conservation}) implies that the sum of the probability
fluxes along the normal direction of the inflow switching surface
is zero. 
\end{enumerate}
Therefore, following the convention of~\cite{Malhame1985}, we will
call~(\ref{eq:Absorbing}) the absorbing condition,~(\ref{eq:Continuity})
the continuity condition, and~(\ref{eq:Conservation}) the probability
conservation condition. In particular, note that for the deterministic
continuous systems where there is no Brownian motion, (i.e., $\Sigma=0$)
the results in Theorems~\ref{thm:MainPDE} and~\ref{thm:MainBC}
still hold.

\begin{rem}
It is worth mentioning that the author in~\cite{Bect2010} derived
a measure-valued formulation of the forward equation for the GSHS
based on the Levy's identity~\cite{Walsh1972,Bass1979}. In contrast,
our method is based on a sufficient characterization of the SHS martingales
(or the Dynkin's formula). This martingale characterization is equivalent
to the Levy's identity in evaluating the expected increments caused
by the jumps. Nevertheless, it extracts more information from the
individual load model, namely the boundary condition satisfied by
the generator of the SHS. The generator boundary condition directly
determines the PDE boundary conditions through the adjoint relation.
However, this information was not employed in~\cite{Bect2010}, and
therefore the results there do not directly apply to the PDE modeling
of responsive load aggregation. In addition, it is worth mentioning
that the presence of the generator boundary condition eliminates the
need of proving the existence of the mean jump intensity as required
by~\cite{Bect2010}.
\end{rem}

\subsection{Load Population with Heterogeneous Parameters~\label{subsec:Cluster}}

In this subsection, we consider the case of heterogeneous parameters.
As shown in (\ref{eq:HS_SDE}), theoretically we can treat the parameter
distribution in the same way as the initial distribution. However, the increase
in dimensions will cause major computational issues. Some of the existing
methods rely on the noise process to account for the parameter heterogeneity
\cite{Moura2013,Moura2014}. Nonetheless, this is generally not an
accurate characterization, since the randomly distributed load parameters
$\theta$ may not be equivalent to an additive drift term to the vector
field $f$ that is normally distributed. Therefore, the corresponding
collective behaviors can be quite different.

Since parameter heterogeneity cannot be captured well by diffusion
terms, we approximate the parameter heterogeneity using several homogeneous
population whose parameters are obtained from clustering over the
samples drawn from the distribution $p_{\theta}$. Denote $p(q,x,t)$
the p.d.f. of the heterogeneous population dynamics~(\ref{eq:HS_SDE}).
Then the conditional p.d.f. $p(q,x,t|\theta^{i})$ represents the
p.d.f. of the homogeneous population with parameters $\theta^{i}$,
that is, the SHS model~(\ref{eq:HS_SDE}) with fixed load parameters
$\theta^{i}$, assuming that it has the same initial p.d.f $p_{0}$
as the heterogeneous population. Intuitively, $p(q,x,t)$ can be estimated
by p.d.f. of sufficiently many homogeneous populations. Specifically,
we have
\begin{prop}
\label{prop:cluster}Let $\theta^{i}$, $i=1,2,\dots n,$ be the samples
drawn from the distribution $p_{\theta}:\Theta\mapsto\mathbb{R}_{+}$,
and assume that the set $\Theta$ are compact. If $p(q,x,t|\theta)$
is continuously dependent on $\theta$, then $\forall q,x,$ and $t>0$,
\begin{equation}
\lim_{n\rightarrow\infty}\frac{1}{n}\overset{{\scriptstyle n}}{\underset{{\scriptstyle i=1}}{\sum}}p(q,x,t|\theta^{i})=p(q,x,t),\,\mbox{almost surely}\label{eq:Sampling}
\end{equation}
\end{prop}
\begin{IEEEproof}
See Appendix~\ref{subsec:ProofProp2}.
\end{IEEEproof}
\begin{rem}
The continuous dependence of the p.d.f. on the initial data may be
most appropriately investigated from the corresponding PDE. However,
it is out of the scope of this paper to go into that direction to
determine the conditions that guarantee the well-posedness of the
PDE. Instead, for the practical problem considered in this paper,
we shall assume this property carried over from the SHS modeling.
Later on, we will demonstrate the effectiveness of our method through
simulation on a realistic example.
\end{rem}
We can further approximate the empirical distribution function $F_{n}$
by $\tilde{F}_{c}\coloneqq\sum_{k=1}^{n_{c}}w_{k}\mathbf{1}_{\{\bar{\theta}^{k}\leq\theta\}}$
using much fewer samples $\bar{\theta}^{k}$, where $n_{c}\ll n$
and $\sum_{k=1}^{n_{c}}w_{k}=1$ such that fixing $n_{c}$, the samples
$\bar{\theta}^{k}$ and the index sets $\mathcal{I}_{k}$ minimize
the sum of the within-cluster distances $\epsilon(n)\coloneqq\sum_{k=1}^{n_{c}}\sum_{i\in\mathcal{I}_{k}}\left\Vert \bar{\theta}^{k}-\theta^{i}\right\Vert $,
where $\theta^{i},\,i\in\{1,2,\dots n\}$ are the original samples.
Hence, $\mathcal{I}_{k}$ contains the indices of $\theta^{i}$ which
are closest to $\bar{\theta}^{k}$, and $w_{k}=\frac{n_{k}}{n},$
$n_{k}=\left|\mathcal{I}_{k}\right|$ is the weight of the $k$th
cluster. In this paper, we adopt the $k-$means clustering algorithm
which efficiently calculates a suboptimal solution to the minimization
of $\epsilon(n)$. Furthermore, by assuming the continuous dependence
on the parameters, the approximation error can be bounded by
\begin{align*}
 & \left|\int p(q,x,t|\theta)dF_{n}-\int p(q,x,t|\theta)d\tilde{F}_{c}\right|\\
\leq & \frac{1}{n}\sum_{k=1}^{n_{c}}\sum_{i\in\mathcal{I}_{k}}\left|p(q,x,t|\bar{\zeta}^{k})-p(q,x,t|\zeta^{i})\right|\\
\leq & \frac{c}{n}\epsilon(n),
\end{align*}
where $c>0$ is a constant depending on the set $\Theta$, and the
specific SDEs. Since the approximation error decreases linearly with
$\epsilon(n)$, we may increase the number of the clusters $n_{c}$
to reduce the error. However, it is a trade-off between the accuracy
and the computational complexity.

\section{Applications\label{sec:App}}

The proposed SHS approach provides a unified framework to obtain aggregate
models of responsive loads. It contains many existing models as special
cases. For example, the pioneer work by~\cite{Malhame1985} can be
easily obtained from Theorems~\ref{thm:MainPDE} and~\ref{thm:MainBC}
(See the following example in Section~\ref{subsec:PDEHVAC}). In
this section, we will discuss three aggregate modeling examples and
use the proposed SHS framework to derive the associated density evolution
PDEs. These examples have not been formally studied in the literature.
Hence, this section not only demonstrates how to use our SHS framework,
but also contains new contributions in the field of aggregate load
modeling.

\subsection{PDE Model for HVACs \label{subsec:PDEHVAC}}

Consider the HVAC aggregation problem similar to Example~\ref{ex: Shvac}.
To draw connections with other results in the literature, we assume
a homogeneous population where each HVAC is modeled by a 2D SDE, 
\[
dx(t)=Ax(t)+B_{q(t)}+\sigma\left[\begin{array}{c}
 dW_{1}(t)\\
dW_{2}(t)
\end{array}\right],
\]
and the same local control rule~(\ref{eq:localcontrol}), where $W_{i}$,
$i=1,2$ are the standard 1D Brownian motions. For the convenience
of comparison, we do not consider the random jump in this example.
The hybrid state space $X$, outflow switching surface $\mathcal{G}$,
and inflow switching surface $\mathcal{S}$ are the same as in~Example~\ref{ex: Shvac},
which is illustrated in Fig.~\ref{fig:HVACss}. We denote the partitions
of $X$ by $X_{q}^{i}=\{q\}\times\Gamma_{i}(q)$, where $\Gamma_{2}(q)\coloneqq\left\{ x\in\mathbb{R}^{2}:u_{\mbox{set}}-\delta<x_{1}<u_{\mbox{set}}+\delta\right\} $
is the overlapped continuous state space, and $\Gamma_{1}(q)$ represents
the rest of the state space given by $X_{q}\backslash\Gamma_{2}(q)$.
Let $p_{i}(q,x,t)$ be the corresponding p.d.f. defined on $\Gamma_{i}(q)$.
By Theorem~\ref{thm:MainPDE}, we can easily have for all $q=0,1$,
$i=1,2$, and $x\in\Gamma_{i}(q)$, 
\begin{multline*}
D_{t}p_{i}(q,x,t)+\nabla\cdot\left(f(q,x)p_{i}(q,x,t)\right)-\frac{\sigma^{2}}{2}\Delta p_{i}(q,x,t)=0,
\end{multline*}
where $f(q,x)=Ax+B_{q}$ and $\Delta$ is the Laplacian operator.

The PDE boundary conditions directly follow from~(\ref{eq:Absorbing})-(\ref{eq:Conservation}).
We have the outer unit normal vector $\nu(0,x)=[1,0]^{T}$ on the
outflow switching surface $\mathcal{G}_{0}$. Similarly, for the ON
mode $\nu(1,x)=[-1,0]^{T}$ on $\mathcal{G}_{1}$. Therefore, the
condition~(\ref{eq:Absorbing}) yields $p_{2}(q,x,t)=0,$ $\forall q=0,1$,
$x\in\partial X_{q}$. Condition~(\ref{eq:Continuity}) reduces to
$p_{1}(q,x,t)=p_{2}(q,x,t)$ on the inflow switching surface $\mathcal{S}$.
Note that by the definition used in Theorem~\ref{thm:MainBC}, the
probability flux $\overrightarrow{\gamma}(q^{+},x,t)$ is defined
in $\Gamma_{2}(q^{+})$, while $\gamma(q,x,t)$ is the outflow probability
flux from $\Gamma_{2}(q)$, for $q=0,1$. Furthermore, using the boundary
conditions obtained above and the continuity of the vector field $f(q,x,t)$,
we have from~(\ref{eq:Conservation}) that for all $q=0,1$ and $x\in\mathcal{G}_{q}$,
the following probability conservation condition holds:
\[
D_{x_{1}}p_{2}(q^{+},x,t)-D_{x_{1}}p_{1}(q^{+},x,t)+D_{x_{1}}p_{2}(q,x,t)=0.
\]

The result in~\cite{Malhame1985} is a simple special case of the
above obtained model (one immediately get the PDE model of~\cite{Malhame1985}
if the $x_{2}$ dynamic is 0). Many other PDE models for aggregating
HVACs~\cite{BashashFathy2013,Moura2013,Moura2014,Ghaffari2014,PaccagnanKamgarpourLygeros2015,Totu2017}
can also be justified or obtained conveniently from Theorems~\ref{thm:MainPDE}
and~\ref{thm:MainBC}.

\subsection{PDE Model for PEV Charging}

We consider the aggregation of PEVs with the charging dynamics as
detailed in Example~\ref{ex: PEVs}. Note that the heterogeneous
parameters such as battery capacity, charging rate, charging deadlines
may all be normalized or translated into the initial distribution
of the timing dynamics. In addition, the spontaneous mode switching
behaviors will be modeled for the realistic consideration. The PDE
model can capture the distribution evolution of the PEVs in different
modes, which will be useful to predict the aggregate power consumption.

By using Theorems~\ref{thm:MainPDE} and~\ref{thm:MainBC},  the
major work to obtain the PDE model is only left to identify the partitions
of the state spaces. We see that only $X_{1}$ is partitioned into
$\Gamma_{11}\coloneqq X_{1}\backslash X_{0}$ and $\Gamma_{12}\coloneqq X_{0}$.
Let $p_{1}(1,x,t)$ and $p_{2}(1,x,t)$ be the p.d.f. on $\Gamma_{11}$
and $\Gamma_{12}$, respectively. Suppose the transition intensity
function $\lambda(q,x)$ is given. Then we obtain directly from~(\ref{eq:FKPDE}):
\begin{flalign*}
D_{t}p(0,x,t)-D_{x_{2}}p(0,x,t)+\lambda(0,x)p(0,x,t) & =0\,\text{in }X_{0},\\
D_{t}p_{1}(1,x,t)-D_{x_{1}}p_{1}(1,x,t)+\lambda(1,x)p_{1}(1,x,t) & =0\,\text{in }\Gamma_{11},\\
D_{t}p_{2}(1,x,t)-D_{x_{1}}p_{2}(1,x,t)+\lambda(1,x)p_{2}(1,x,t)\\
-\lambda(0,x)p(0,x,t) & =0\,\text{in }\Gamma_{12},\\
D_{t}p(2,x,t)-D_{x_{1}}p_{2}(1,x,t) & =0\,\text{in }X_{2}.
\end{flalign*}

Since $\Sigma$ is a zero matrix, there are only probability conservation
conditions. On the outflow switching surface $\mathcal{G}_{0}$, we
have $\nu(0,x)=[0,-1]^{T}$. Let $S_{1}=\{x_{1}=0,x_{2}<0\}$ and
$S_{2}=\{x_{1}=0,x_{2}\geq0\}$ be the switching surfaces of $\Gamma_{12}$
and $\Gamma_{11}$, respectively, on both of which we have $\nu(1,x)=[-1,0]^{T}.$
Then we can obtain from (\ref{eq:Conservation}) the following PDE
boundary conditions: 
\begin{flalign*}
p(0,x,t) & =0,\,\forall x\in\mathcal{G}_{0},\\
p(2,x,t) & =p_{h}(1,x,t),\,\forall x\in S_{h},\,h=1,2.
\end{flalign*}

\subsection{Price Response of Aggregate HVACs}

In this example, we consider the price response of HVACs with heterogeneous
parameters $\theta=(A,\,B_{1},B_{2},\,u_{\mbox{set}},\,\delta)$,
see~(\ref{eq:HVACsETP}) in Example~\ref{ex: Shvac} for the 2D
hybrid system model of the individual HVAC load. The price responsive
HVACs can adjust their power state according to a price signal. In
addition to the local control rules~(\ref{eq:localcontrol}), the
setpoint of the HVAC can be modified by an external price signal $\upsilon(t)$
according to a adjustable price response curve~\cite{Li2014}. We
will derive the PDE model for a heterogeneous population of HVACs,
and use it to characterize the aggregate power response under the
price variation. Note that different from Section~\ref{subsec:PDEHVAC},
we will not consider the noise process in this example.

As discussed in Section~\ref{subsec:Cluster}, we can cluster the
heterogeneous parameters $\theta$ to obtain several homogeneous populations.
Then the p.d.f. of the heterogeneous HVACs can be approximated by
the weighted sum of the solutions of the PDE models with homogeneous
parameters. Specifically, the coefficients of each PDEs are evaluated
at the cluster centers of the sampled parameters. For example, the
p.d.f. evolution of the $k$th cluster is governed by:
\begin{equation}
D_{t}p_{i}^{k}(q,x,t)+\nabla\cdot\left(f^{k}(q,x)p_{i}^{k}(q,x,t)\right)=0,\label{eq:PDEcluster}
\end{equation}
in $\Gamma_{i}(q)$, $\forall q=0,1$, $i=1,2$, (see Subsection \ref{subsec:PDEHVAC}
for the definition of $\Gamma_{i}(q)$), where $f^{k}(q,x)$ is the
vector field at the $k$th center, $k=1,2,...,n_{c}$. The boundary
conditions are obtained from Theorem~\ref{thm:MainBC} as 
\begin{multline*}
f^{k}(q^{+},x)p_{1}^{k}(q^{+},x,t)-f^{k}(q^{+},x)p_{2}^{k}(q^{+},x,t)\\
-f^{k}(q,x)p_{2}^{k}(q,x,t)=0,
\end{multline*}
for $x=u_{\mbox{set}}+\delta$ if $q=0$, and $x=u_{\mbox{set}}-\delta$
if $q=1$.

The above PDE and boundary conditions can be easily obtained from
the model derived in Section~\ref{subsec:PDEHVAC} by setting $\sigma=0$.
Once $p^{k}(q,x,t)$ is obtained for each cluster, the aggregate power
response can be calculated as 
\[
y^{a}(t)=NW\overset{{\scriptstyle n_{c}}}{\underset{{\scriptstyle k=1}}{\sum}}w_{k}\int_{X_{1}}p^{k}(1,x,t)dx,
\]
where $N$ is the total number of the HVACs , $W$ is the power consumption
of one HVAC, and $w_{k}$ is the weight of the $k$th cluster.

Without loss of generality, we assume that the price changes at $t=3$
hour and $t=6$ hour, which cause the setpoints changes of all HVACs
by $1^{\circ}\mbox{F}$ and $-1^{\circ}\mbox{F}$, respectively. For
the Monte Carlo simulation, the load parameters are generated randomly
using GridLAB-D \cite{GridLAB}, and then the ETP models are derived
using these parameters. The initial setpoints for the 2000 HVACs are
generated uniformly between $[70^{\circ}\mbox{F},78^{\circ}\mbox{F}]$,
and the initial temperatures for air and internal solid mass are generated
uniformly within the temperature deadband. For the PDE simulation,
the $k$-means clustering method is performed over the system matrices
(which are generated by the load parameter and note that $f$ is linear
in $x$ and $q$) to obtain 10 clusters. The initial distribution
$p_{\xi_{0}}$ is given arbitrarily since the aggregate power output
converges to the steady state very fast before the setpoint changes.
Then the PDEs in (\ref{eq:PDEcluster}) are solved for each cluster
using the Donor-Cell finite volume scheme along with the dimensional
splitting method for multi-dimensional problem \cite{LeVeque2002}.
We compare in Fig. \ref{fig:MCvsWeightedPDE} the aggregate response
of Monte Carlo simulation of 2000 HVACs and that of the weighted PDEs
of 10 clusters. Clearly, the weighted PDE model captures the major
oscillation cycles after the price/setpoint changes very accurately.
\begin{center}
\begin{figure}[t]
\centering{}\includegraphics[width=0.9\linewidth]{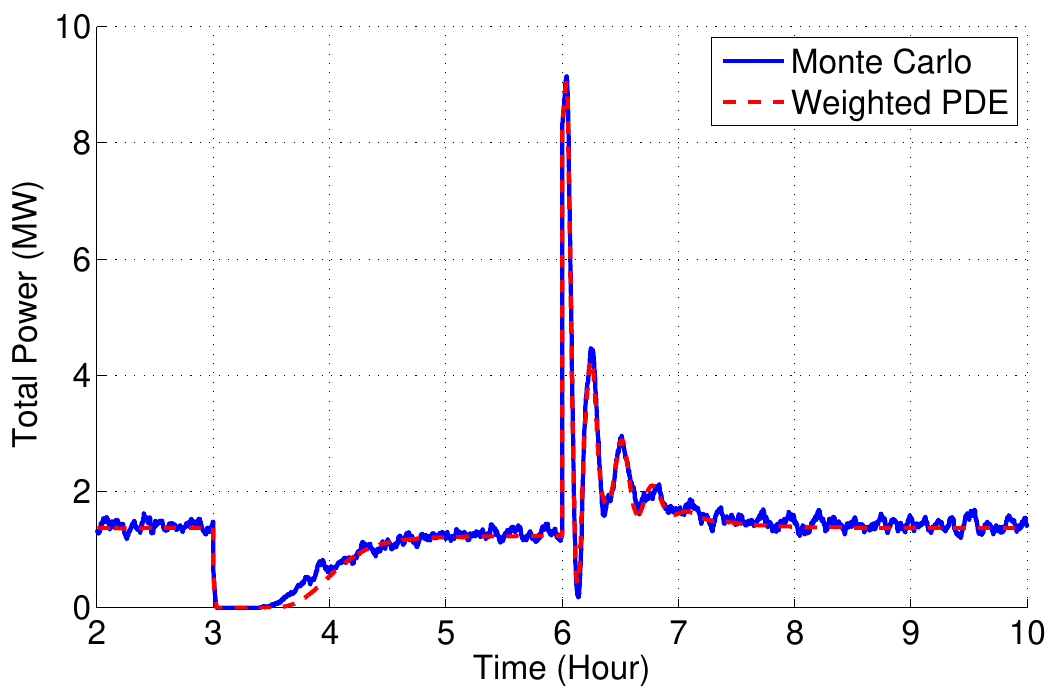}
\caption{\label{fig:MCvsWeightedPDE}Comparison of aggregate response between
Monte Carlo simulation and weighted PDE using 10 clusters.}
\end{figure}
\par\end{center}

\section{Conclusion}

This paper developed a unified stochastic hybrid system (SHS) framework
for the aggregation of a large population of responsive loads. General
nonlinear stochastic differential equations were used to describe
the continuous state evolution between discrete mode transitions,
and the mode transition was modeled by both deterministic and random
jumps. The PDE satisfied by the hybrid-state probability density function
was derived based on the adjoint relationship with the generator of
the SHS process. In particular, the PDE boundary conditions were uniquely
determined from the generator boundary condition. These results generalized
many existing models and can be directly applied to more general modeling
scenarios. Several examples were provided to illustrate the effectiveness
of the proposed modeling framework.

\appendix{}

\subsection{Proof of Lemma~\ref{lem:L1Mart}\label{subsec:Proof1} (Martingale
Characterization)}

Following the same steps in \cite[Proof of Theorem 2]{Bujorianu2006},
we can represent $C_{t}^{\psi}$ as 
\begin{equation}
C_{t}^{\psi}=\int_{0}^{t}\nabla\psi(\xi_{t})\cdot\sigma(\xi_{t})dW(t)+\int_{[0,t]\times X}B\psi dq,\label{eq:CtfMar}
\end{equation}
where $q(t,A)$, $\forall A\in\mathcal{B}(X)$ is the stochastic process
defined in \cite[page 67, Proposition 26.7]{Davis1993a}. We claim
that both terms on the right side of (\ref{eq:CtfMar}) are martingales
with respect to the natural filtration $\mathcal{F}_{t}$ generated
by the GSHS process $\xi_{t}$. Note that all the positive constants
will be denoted by $c$ in the following. 

To show that the first term is a $\mathcal{F}_{t}$-martingale, we
will prove that all the conditions of~\cite[page 33, Corollary 3.2.6]{Oksendal2003}
hold. We know $\sigma_{ij}^{q}\in C_{b}^{1}$ by Assumption~\ref{Ass:regularity}.
Therefore, we can have 
\[
\vert\sigma_{ij}(x)\vert\leq c(1+\left\Vert x\right\Vert ),
\]
for some constant $c>0$. Since $\psi^{q}\in C_{b}^{1}$, we have
$\nabla\psi$ is bounded. Note that the first term can be written
as 
\[
\int_{0}^{t}\sum_{j=1}^{m}\sum_{i=1}^{n}(\nabla\psi)_{i}\sigma_{ij}dW_{j}(t),
\]
where $(\nabla\psi)_{i}$ and $W_{i}(t)$ are the $i$th entry of
$\nabla\psi$ and $W(t)$, respectively. Then for a fixed index $j$,
$\forall t\geq0$, we have {\small{}
\begin{align}
E\left[\int_{0}^{t}\left(\sum_{i=1}^{n}(\nabla\psi)_{i}\sigma_{ij}\right)^{2}dt\right] & \leq cE\left[\int_{0}^{t}(1+\left\Vert x\right\Vert )^{2}dt\right]<\infty,\label{eq:Wmartingale}
\end{align}
}where the last inequality is by Remark~\ref{rem:Remark1}. Clearly,
$\nabla\psi(\xi_{t})\sigma(\xi_{t})$ is $\mathcal{F}_{t}$-adapted.
Together with the fact that $\xi_{t}$ is c{\`{a}}dl{\`{a}}g~\cite{Bujorianu2006},
it can be concluded that $\nabla\psi\sigma$ as a function of $(t,\omega)\in[0,\infty)\times\Omega$
is $\mathcal{B}\times\mathcal{F}$ measurable~\cite[page 5, Proposition 1.13]{Karatzas1991},
where $\mathcal{B}$ is the Borel $\sigma$-algebra on $[0,\infty)$.
Combining the above results, it can be concluded that $\int_{0}^{t}\nabla\psi\sigma dW(t)$
is an $\mathcal{F}_{t}$-martingale by the standard results of stochastic
integration, see for example~\cite[Corollary 3.2.6]{Oksendal2003}.

For the second term, the proof is similar to~\cite[page 272, Theorem (A4.9)]{Davis1993a}
for the single jump process and~combines the idea of~\cite[Lemma 3.1]{Elliott1976}
to decompose a general jump process into the summation of single jump
processes. However, the natural filtration and the probability measure
$P(\omega)$ of the GSHS are different from those of the general jump
process due to the additional It{\^{o}}  diffusions between jumps.
To prove the claim, we can express $P(\omega)$ explicitly as follows,
for $t\in[T_{k},T_{k+1})$,
\begin{multline}
\int_{\Omega}\psi(\xi_{t}(\omega))dP(\omega)=\\
\int_{\Omega_{0}}\int_{\Omega_{1}}\cdots\int_{\Omega_{k}}\psi(\xi_{t}(\omega))dP_{\xi(T_{k})}^{k}(\omega^{k})\mu^{k}(dY_{k};\omega^{k-1})\cdots\\
\times dP_{\xi(T_{1})}^{1}(\omega^{1})\mu^{1}(dY_{1};\omega^{0})dP_{\xi_{0}}^{0}(\omega^{0})\mu^{0}(d\xi_{0}),\label{eq:GSHSProb}
\end{multline}
where $\omega$ is the concatenation of the trajectories of the It{\^{o}}
 diffusions $\omega^{k}\in\Omega_{k}$ killed at time $T_{k+1}-T_{k}$
with $\omega^{k}(0)=\omega(T_{k})=\xi_{T_{k}}$, $\forall k\in\mathbb{N}$.
The measure $P_{\xi(T_{k})}^{k}$ is the Wiener probability corresponding
to the $k$th It{\^{o}}  diffusion starting at $Y_{k}\coloneqq(\xi_{T_{k}},T_{k})$.
The measure of the post-jump position and the jump time of the $k$th
jump conditioned on the previous It{\^{o}} diffusion $\omega^{k-1}$
is defined by $\forall A\in\mathcal{B}(X)$, $\forall s>T_{k-1}$,
\[
\mu^{k}(A\times(s-T_{k-1},\infty);\omega^{k-1})\coloneqq\mathcal{R}(\xi_{s^{-}},A)\beta(s-T_{k-1};\omega^{k-1}),
\]
where $\mathcal{R}$ is the transition measure defined in Lemma~\ref{lem:BM}
and $\beta$ is the survival function defined in~(\ref{eq:Survival}).
Thus using~(\ref{eq:GSHSProb}), several intermediate results in
proving~\cite[Theorem (A4.9)]{Davis1993a} and~\cite[Lemma 3.1]{Elliott1976}
can be easily verified by direct calculations. Hence, in the following
we will leave out the explicit calculations, and instead, focus on
the key arguments of the proof.

Following~\cite[Lemma 3.1]{Elliott1976}, we can decompose $q(t,A)$
into the sum of single jump processes conditioned on their immediate
precedent It{\^{o}} diffusions, that is, $q(t,A)=\sum_{k\geq1}q^{k-1}(t,A;\omega^{k-1})$.
Then by direct calculation using (\ref{eq:GSHSProb}) in a similar
way to~\cite[Theorem (A4.4)]{Davis1993a}, it can be shown that each
$q^{k-1}$ is an $\mathcal{F}_{t}$-martingales (which is often called
martingale-valued measures or simply martingale measures~\cite[page 105]{Applebaum2009}).
Since $B\psi$ is bounded, it can be proved similarly that $\int_{[0,t]\times X}B\psi dq^{k-1}$
is also an $\mathcal{F}_{t}$-martingales. It follows immediately
that $q(t\land T_{k},A)$ is a uniformly integrable $\mathcal{F}_{t}$-martingale
for each $k\in\mathbb{Z}_{+}$, and so is $\int_{[0,t\land T_{k}]\times X}B\psi dq$.
Now the result in~\cite[Proposition 1.7]{Revuz1999} implies that
in order to show $\int_{[0,t]\times X}B\psi dq$ is a $\mathcal{F}_{t}$-martingale,
it suffices to prove that $B\psi\mathbf{1}_{\{s\leq t\}}\in L_{1}(q)$
for all $t\geq0$. Through a similar calculation to~\cite[Theorem (A4.5)]{Davis1993a},
this condition can be shown to be equivalent to $B\psi\mathbf{1}_{\{s\leq t\}}\in L_{1}(\tau)$
for all $t\geq0$, where $\tau$ is a counting process defined earlier
in (\ref{eq:countingP}). The latter condition can be further evaluated
as $\forall t\geq0$,
\begin{multline}
E_{\xi_{0}}\left(\int\left|B\psi\mathbf{1}_{\{s\leq t\}}\right|d\tau\right)=\\
E_{\xi_{0}}\left(\sum_{k=1}^{N_{t}}\left|\psi(\xi_{T_{k}})-\psi(\xi_{T_{k}^{-}})\right|\right),\label{eq:BfLoc}
\end{multline}
which is finite since $\psi$ is bounded and $E_{\xi_{0}}(N_{t})<\infty$
by Remark \ref{rem:Nt}. Hence, $B\psi\mathbf{1}_{\{s\leq t\}}\in L_{1}(\tau)$
and we conclude that the second term on the right side of~(\ref{eq:CtfMar})
is a $\mathcal{F}_{t}$-martingale. This completes the proof of this
lemma. \hfill \qed

\subsection{Proof of Proposition~\ref{prop:Lf}\label{subsec:Proofcor1} (Extended
Generator)}

The formula in (\ref{eq:Generator}) is obtained by taking the transition
measure $\mathcal{R}$ in Lemma~\ref{lem:BM} as a Dirac measure
(unit mass), $\forall(q,x)\in\bar{X}$,
\begin{equation}
\delta_{(q^{+},x)}(\zeta;(q,x))=\begin{cases}
1, & \mbox{if }\zeta=(q^{+},x);\\
0, & \mbox{otherwise}.
\end{cases}\label{eq:transition}
\end{equation}

Similarly, by using (\ref{eq:transition}), the boundary condition
in Lemma~\ref{lem:BM} reduces to $\psi(q,x)=\psi(q^{+},x),$ $\forall(q,x)\in\mathcal{G}$.\hfill \qed

\subsection{Proof of Proposition~\ref{prop:DynkinCt}\label{subsec:ProofCorDynkinCt}
(Dynkin's Formula)}

This directly follows from Corollary~\ref{prop:Lf} and Lemma~\ref{lem:L1Mart}.
The boundedness of $\psi$ guarantees that $B\psi\in L_{1}^{\mbox{loc}}(\tau)$
as can be seen from~(\ref{eq:BfLoc}) in the proof of Lemma~\ref{lem:L1Mart}.
Therefore, the set $\mathcal{D}$ satisfies all the conditions for
$\psi$ in Corollary~\ref{prop:Lf} and Lemma~\ref{lem:L1Mart}.
Hence, $C_{t}^{\psi}$ is a martingale. The Dynkin's formula follows
by taking the expectation of $C_{t}^{\psi}$ and noticing that it
is~$0$.\hfill \qed

\subsection{Proof of Theorem~\ref{thm:MainPDE}\label{subsec:ProofThm1} (PDE
Characterization)}

The following divergence theorem simplifies our main proofs of obtaining
the forward equation. It does not require the vector field $g$ to
be compactly supported on $M$. Although the original theorem given
in~\cite{Driver2002} is stated for the manifold with boundary, the
same result holds for the manifold with corners and can be proved
in a similar way.
\begin{lem}
\label{lem:divergence}(Divergence Theorem) Let $M\subset\mathbb{R}^{n}$
be an oriented $C^{2}$-manifold with corners and $M$ is closed,
and $\nu:\ \partial M\mapsto\mathbb{R}^{n}$ be the unit outward pointing
normal to $M$. If $g:M\mapsto\mathbb{R}^{n}$ is continuous on $M$
and $C^{1}$ in $M^{\circ}$, and 
\begin{equation}
\int_{M}\left\{ \left|g\right|+\left|\emph{\ensuremath{\nabla}}\cdot g\right|\right\} dV+\int_{\partial M}\left|g\cdot\nu\right|dS<\infty,\label{eq:IntCon}
\end{equation}
then 
\[
\int_{\partial M}g\cdot\nu dS=\int_{M}\emph{\ensuremath{\nabla}}\cdot gdV,
\]
where $S$ is the surface measure on $\partial M$ and $V$ is the
volume measure.
\end{lem}
\begin{IEEEproof}[Proof of Theorem~\ref{thm:MainPDE}]
By Proposition~\ref{prop:DynkinCt}, $\forall\psi\in\mathcal{D}$,
the Dynkin's formula (\ref{eq:Dynkin}) holds. Evaluate the expectation
using $p(q,x,t)$, and take the time derivative of both sides of (\ref{eq:Dynkin}),
yielding
\begin{equation}
\frac{\partial}{\partial t}\underset{{\scriptstyle q}}{\sum}\int_{X_{q}}\psi(q,x)p(q,x,t)dV=E\left[\hat{L}\psi(q(t),x(t))\right],\label{eq:dt}
\end{equation}

By the bounded convergence theorem, we can exchange the time derivative
and the integral, and it follows that
\begin{multline}
\underset{{\scriptstyle q}}{\sum}\int_{X_{q}}\biggl[\psi(q,x)\frac{\partial p(q,x,t)}{\partial t}-\emph{\ensuremath{\nabla}}\psi(q,x)\cdot f(q,x)p(q,x,t)\\
-\frac{1}{2}Tr\left(\Sigma(q,x)\emph{\ensuremath{\nabla}}^{2}\psi(q,x)\right)p(q,x,t)\\
-\lambda(q,x)\left(\psi(q^{+},x)-\psi(q,x)\right)p(q,x,t)\biggr]dV=0,\label{eq:PreDiV}
\end{multline}

By the chain rule of differentiation, the second term in (\ref{eq:PreDiV})
can be written as
\begin{align}
\emph{\ensuremath{\nabla}}\psi\cdot fp & =\mbox{\emph{\ensuremath{\nabla}}}\cdot\left(fp\psi\right)-\psi\mbox{\emph{\ensuremath{\nabla}}}\cdot\left(fp\right),\label{eq:advection}
\end{align}
and the third term can be written as
\begin{eqnarray}
\frac{1}{2}Tr\left(\Sigma\emph{\ensuremath{\nabla}}^{2}\psi\right)p & = & \frac{1}{2}\underset{{\scriptstyle j=1}}{\overset{{\scriptstyle n}}{\sum}}p\Sigma_{j}\left(\emph{\ensuremath{\nabla}}\psi\right)_{x_{j}}\nonumber \\
 & = & \frac{1}{2}\underset{{\scriptstyle j=1}}{\overset{{\scriptstyle n}}{\sum}}\left(p\Sigma_{j}\emph{\ensuremath{\nabla}}\psi\right)_{x_{j}}-\left(p\Sigma_{j}\right)_{x_{j}}\emph{\ensuremath{\nabla}}\psi\nonumber \\
 & = & \frac{1}{2}\emph{\ensuremath{\nabla}}\cdot\left(p\Sigma\emph{\ensuremath{\nabla}}\psi\right)-\frac{1}{2}\emph{\ensuremath{\nabla}}\cdot\left(\emph{\ensuremath{\nabla}}\cdot(p\Sigma)\psi\right)\nonumber \\
 &  & +\frac{1}{2}\emph{\ensuremath{\nabla}}\cdot\left(\emph{\ensuremath{\nabla}}\cdot(p\Sigma)\right)\psi.\label{eq:DiffusionTerm}
\end{eqnarray}
where we dropped the dependence on the time and state where no confusion
arises. 

Suppose that all the vector fields under the divergence operation
in (\ref{eq:advection}) and (\ref{eq:DiffusionTerm}) satisfy the
absolutely integrable condition (\ref{eq:IntCon}), and then we can
use Lemma \ref{lem:divergence} in (\ref{eq:PreDiV}) to shift the
differential operator from $\psi$ to $p$. Since $p$ is piecewise
$C^{2}$ in the interior of each $X_{q}$, we can apply the divergence
theorem to each partition of $X$. It then follows after grouping
the terms that
\begin{equation}
\underset{{\scriptstyle q,i}}{\sum}\int_{X_{q}^{i}}T[p]\psi dV-\int_{\partial X_{q}^{i}}\left(\gamma\psi+\frac{1}{2}p\Sigma\emph{\ensuremath{\nabla}}\psi\right)\cdot\nu(i)dS=0,\label{eq:component}
\end{equation}
where $T[p]$ and $\gamma$ is defined in~(\ref{eq:FKPDE}) and $\nu(i)$
is the outer unit normal vector on $\partial X_{q}^{i}$. Now let
$\psi$ be compactly supported in one of the partition $X_{q}^{i}$.
Then all the surface integrals in (\ref{eq:component}) vanishes,
and we are left with 
\[
\int_{X_{q}^{i}}T[p]\psi dV=0.
\]

Since $\psi$ is arbitrary, we must have $T[p]=0$. Thus it completes
the proof.
\end{IEEEproof}

\subsection{Proof of Theorem~\ref{thm:MainBC}\label{subsec:ProofThm2} (Boundary
Conditions)}

Note that in general the surface integral in~(\ref{eq:component})
will be integrated over $\partial X$ once, and over $\mathcal{E}$
twice from both sides, where recall that $\partial X=\cup_{q}\{q\}\times\partial X_{q}$
and $\mathcal{E}=\cup_{q}\{q^{+}\}\times\partial X_{q}$. We first
claim that it is equivalent to evaluate the surface integral in~(\ref{eq:component})
over $\mathcal{G}$ and both sides of $\mathcal{S}$ only. Since
$\psi^{q}\in C^{2}$, $\forall q\in Q$, $\psi$ and $\emph{\ensuremath{\nabla}}\psi$
are continuous over $\mathcal{E}$\textbackslash{}$\mathcal{S}$.
Hence, the surface integrals on both sides of $\mathcal{E}$\textbackslash{}$\mathcal{S}$
cancel each other out. By the definition of $\mathcal{G}_{q}$ in~(\ref{eq:GqSHS}),
we have $\nu\cdot f\leq0$ and $\nu^{T}\Sigma=\boldsymbol{0}$ on
$\partial X\backslash\mathcal{G}$. If $\nu\cdot f=0$ and $\nu^{T}\Sigma=\boldsymbol{0}$,
then clearly the integral is $0$ over$\partial X\backslash\mathcal{G}$.
If $\nu\cdot f<0$ and $\nu^{T}\Sigma=\boldsymbol{0}$, we have two
situations: 1) on $(\partial X\backslash\mathcal{G})\backslash\mathcal{S}$,
we must have $p(q,x,t)=0$ for all $t>0$ (since all particles left
the boundary immediately after the starting time and no particle jumps
to the boundary from other modes), and hence, the surface integral
is $0$ on $(\partial X\backslash\mathcal{G})\backslash\mathcal{S}$;
2) on $(\partial X\backslash\mathcal{G})\cap\mathcal{S}$: the integration
over this set is in fact included in the integration over $\mathcal{E}$.
Thus it proves the claim.

Now using the generator boundary condition $\psi(q,x)=\psi(q^{+},x),$
$\forall(q,x)\in\mathcal{G}_{q}$ in Proposition~\ref{prop:Lf},
we can collect the $\gamma\psi$ terms which are integrated on $\mathcal{G}$
and on both sides of $\mathcal{S}$. Furthermore, since $\emph{\ensuremath{\nabla}}\psi$
is continuous, the $\frac{1}{2}p\Sigma\emph{\ensuremath{\nabla}}\psi$
terms integrated on both sides of $\mathcal{S}$ can be collected.
The last group is the $\frac{1}{2}p\Sigma\emph{\ensuremath{\nabla}}\psi$
integrated on $\mathcal{G}$. Moreover, noting that the outer normal
directions of $\mathcal{G}_{q}$ as in $\mathcal{G}$ and in $\mathcal{S}$
are the same, we can obtain from~(\ref{eq:component}) that 
\begin{multline}
\underset{{\scriptstyle q,i}}{\sum}\int_{X_{q}^{i}}T[p]\psi(q,x)dV-\underset{{\scriptstyle q}}{\sum}\left(\int_{\mathcal{G}_{q}}F\psi(q^{+},x)dS\right.\\
\left.+\int_{\mathcal{G}_{q}}H\emph{\ensuremath{\nabla}}\psi(q^{+},x)dS+\int_{\mathcal{G}_{q}}G\emph{\ensuremath{\nabla}}\psi(q,x)dS\right)=0,\label{eq:AfterDiv}
\end{multline}
where $F$, $H$, $G$ denote the left sides of~(\ref{eq:Conservation}),
(\ref{eq:Continuity}) and~(\ref{eq:Absorbing}), respectively.

From Theorem~\ref{thm:MainPDE} we know that the volume integral
in~(\ref{eq:AfterDiv}) is 0. Since $\psi(q^{+},x)$, $\emph{\ensuremath{\nabla}}\psi(q,x)$,
and $\emph{\ensuremath{\nabla}}\psi(q^{+},x)$ can be arbitrary continuous
functions on $\mathcal{G}_{q}$, then $F$, $G$, and $H$ must vanish
on $\mathcal{G}_{q}$, which are the boundary conditions in~(\ref{eq:Absorbing})-(\ref{eq:Conservation}).\hfill \qed

\subsection{Proof of Proposition~\ref{prop:cluster}\label{subsec:ProofProp2}
(Clustering)}

The density functions for the heterogeneous and the homogeneous populations
can be related by
\begin{align*}
p(q,x,t) & =\int_{\Theta}p(q,x,t|\theta)p_{\theta}d\theta.
\end{align*}

Let $F_{\theta}$ be the cumulative distribution function of $\theta$
such that $dF_{\theta}=p_{\theta}d\theta$. Let $F_{n}(r):=\frac{1}{n}\sum_{i=1}^{n}\mathbf{1}_{\{\theta^{i}\leq r\}}$
be the corresponding empirical distribution function with independent
and identically distributed $(\theta^{i})\sim p_{\theta}$, $i\leq n$.
Then by the Glivenko-Cantelli theorem~\cite[Theorem 2.4.7]{Durrett2010},
as $n$ goes to infinity, 
\[
\sup_{r\in\Theta}|F_{n}(r)-F_{\theta}(r)|\rightarrow0,
\]
almost surely.

Since $p(q,x,t|\theta)$ is continuously dependent on $\theta$, it
is bounded on the compact sets $\Theta$. Then it follows from the
dominated convergence theorem that, as $n$ goes to infinity, $\int p(q,x,t|\theta)dF_{n}\rightarrow p(q,x,t)$
a.s., which is exactly~(\ref{eq:Sampling}).\hfill \qed


\begin{thebibliography}{10}
\providecommand{\url}[1]{#1}
\csname url@samestyle\endcsname
\providecommand{\newblock}{\relax}
\providecommand{\bibinfo}[2]{#2}
\providecommand{\BIBentrySTDinterwordspacing}{\spaceskip=0pt\relax}
\providecommand{\BIBentryALTinterwordstretchfactor}{4}
\providecommand{\BIBentryALTinterwordspacing}{\spaceskip=\fontdimen2\font plus
\BIBentryALTinterwordstretchfactor\fontdimen3\font minus
  \fontdimen4\font\relax}
\providecommand{\BIBforeignlanguage}[2]{{%
\expandafter\ifx\csname l@#1\endcsname\relax
\typeout{** WARNING: IEEEtran.bst: No hyphenation pattern has been}%
\typeout{** loaded for the language `#1'. Using the pattern for}%
\typeout{** the default language instead.}%
\else
\language=\csname l@#1\endcsname
\fi
#2}}
\providecommand{\BIBdecl}{\relax}
\BIBdecl

\bibitem{Callaway2009}
D.~S. Callaway, ``Tapping the energy storage potential in electric loads to
  deliver load following and regulation with application to wind energy,''
  \emph{Energy Conversion and Management}, vol.~50, no.~5, pp. 1389 --1400, May
  2009.

\bibitem{Koch2011}
S.~Koch, J.~L. Mathieu, and D.~S. Callaway, ``Modeling and control of
  aggregated heterogeneous thermostatically controlled loads for ancillary
  services,'' in \emph{17th Power Systems Computation Conference}, Stockholm,
  Sweden, August 2011.

\bibitem{Mathieu2013}
J.~L. Mathieu, S.~Koch, and D.~Callaway, ``State estimation and control of
  electric loads to manage real-time energy imbalance,'' \emph{IEEE
  Transactions on Power Systems}, vol.~28, no.~1, pp. 430--440, Feb. 2013.

\bibitem{Zhang2013}
W.~Zhang, J.~Lian, C.-Y. Chang, and K.~Kalsi, ``Aggregated modeling and control
  of air conditioning loads for demand response,'' \emph{IEEE Transactions on
  Power Systems}, vol.~28, no.~4, pp. 4655--4664, Nov. 2013.

\bibitem{MathieuKamgarpourLygerosEtAl2015}
J.~L. Mathieu, M.~Kamgarpour, J.~Lygeros, G.~Andersson, and D.~S. Callaway,
  ``Arbitraging intraday wholesale energy market prices with aggregations of
  thermostatic loads,'' \emph{IEEE Transactions on Power Systems}, vol.~30,
  no.~2, pp. 763--772, March 2015.

\bibitem{li2016market}
S.~Li, W.~Zhang, J.~Lian, and K.~Kalsi, ``Market-based coordination of
  thermostatically controlled loads-part i: A mechanism design formulation,''
  \emph{IEEE Transactions on Power Systems}, vol.~31, no.~2, pp. 1170--1178,
  2016.

\bibitem{TPS15_Part2}
------, ``Market-based coordination of thermostatically controlled loads-part
  ii: Unknown parameters and case studies,'' \emph{IEEE Transactions on Power
  Systems}, vol.~31, no.~2, pp. 1179 -- 1187, 2016.

\bibitem{Malhame1985}
R.~Malham{\'{e}} and C.-Y. Chong, ``Electric load model synthesis by diffusion
  approximation of a high-order hybrid-state stochastic system,'' \emph{IEEE
  Transactions on Automatic Control}, vol.~30, no.~9, pp. 854--860, Sept. 1985.

\bibitem{BashashFathy2013}
S.~Bashash and H.~K. Fathy, ``Modeling and control of aggregate air
  conditioning loads for robust renewable power management,'' \emph{IEEE
  Transactions on Control Systems Technology}, vol.~21, no.~4, pp. 1318--1327,
  July 2013.

\bibitem{Moura2013}
S.~Moura, V.~Ruiz, and J.~Bendtsen, ``Modeling heterogeneous populations of
  thermostatically controlled loads using diffusion-advection {PDE}s,'' in
  \emph{ASME Dynamic Systems and Control Conference, Stanford, CA}, 2013.

\bibitem{Moura2014}
S.~Moura, J.~Bendtsen, and V.~Ruiz, ``Parameter identification of aggregated
  thermostatically controlled loads for smart grids using {PDE} techniques,''
  \emph{International Journal of Control}, vol.~87, no.~7, pp. 1373--1386,
  2014.

\bibitem{Ghaffari2014}
A.~Ghaffari, S.~Moura, and M.~Krsti{\'{c}}, ``Analytic modeling and integral
  control of heterogeneous thermostatically controlled load populations,'' in
  \emph{ASME Dynamic Systems and Control Conference, San Antonio, Texas}, 2014.

\bibitem{liu2016model}
M.~Liu and Y.~Shi, ``Model predictive control of aggregated heterogeneous
  second-order thermostatically controlled loads for ancillary services,''
  \emph{IEEE Transactions on Power Systems}, vol.~31, no.~3, pp. 1963--1971,
  2016.

\bibitem{Davis1993a}
M.~H.~A. Davis, \emph{Markov Models and Optimization}, ser. Monographs on
  Statistics and Applied Probability.\hskip 1em plus 0.5em minus 0.4em\relax
  Chapman \& Hall, 1993, vol.~49.

\bibitem{Bujorianu2012}
L.~M. Bujorianu, \emph{Stochastic Reachability Analysis of Hybrid
  Systems}.\hskip 1em plus 0.5em minus 0.4em\relax Springer, 2012.

\bibitem{HuJ2000}
J.~Hu, J.~Lygeros, and S.~Sastry, ``Towards a theory of stochastic hybrid
  systems,'' in \emph{Lecture Notes in Computer Science}, N.~Lynch and
  B.~Krogh, Eds.\hskip 1em plus 0.5em minus 0.4em\relax Springer Berlin
  Heidelberg, 2000, vol. 1790, pp. 160--173.

\bibitem{Blom2003}
H.~A. Blom, ``Stochastic hybrid processes with hybrid jumps,'' in
  \emph{Analysis and Design of Hybrid System}.\hskip 1em plus 0.5em minus
  0.4em\relax IFAC Press, 2003, pp. 319--324.

\bibitem{Yin2010}
G.~G. Yin and C.~Zhu, \emph{Hybrid Switching Diffusions: Properties and
  Applications}.\hskip 1em plus 0.5em minus 0.4em\relax Springer-Verlag New
  York, 2010.

\bibitem{Bujorianu2006}
M.~L. Bujorianu and J.~Lygeros, ``Toward a general theory of stochastic hybrid
  systems,'' in \emph{Lecture Notes in Control and Information Science},
  H.~Blom and J.~Lygeros, Eds.\hskip 1em plus 0.5em minus 0.4em\relax Springer
  Berlin Heidelberg, 2006, vol. 337, pp. 3--30.

\bibitem{Hespanha2005}
J.~P. Hespanha, ``A model for stochastic hybrid systems with application to
  communication networks,'' \emph{Nonlinear Analysis: Theory, Methods \&
  Applications}, vol.~62, no.~8, pp. 1353--1383, Sept. 2005.

\bibitem{Bect2010}
J.~Bect, ``{A unifying formulation of the Fokker-Planck-Kolmogorov equation for
  general stochastic hybrid systems},'' \emph{Nonlinear Analysis: Hybrid
  Systems}, vol.~4, no.~2, pp. 357--370, May 2010.

\bibitem{Walsh1972}
J.~B. Walsh and M.~Weil, ``\BIBforeignlanguage{fre}{Repr{\'{e}}sentation de
  temps terminaux et applications aux fonctionnelles additives et aux
  syst\`{e}mes de l{\'{e}}vy},'' \emph{\BIBforeignlanguage{fre}{Annales
  scientifiques de l'{\'{E}}cole Normale Sup{\'{e}}rieure}}, vol.~5, no.~1, pp.
  121--155, 1972.

\bibitem{Bass1979}
R.~F. Bass, ``Adding and subtracting jumps from markov processes,''
  \emph{Transactions of the American Mathematical Society}, vol. 255, pp.
  363--376, Nov. 1979.

\bibitem{Wilson1985}
N.~W. Wilson, B.~S. Wagner, and W.~G. Colborne, ``Equivalent thermal parameters
  for an occupied gas-heated house,'' \emph{ASHRAE Transactions}, vol.~91, no.
  CONF-850606-, 1985.

\bibitem{Kalsi2011}
K.~Kalsi, F.~Chassin, and D.~Chassin, ``{Aggregated modeling of thermostatic
  loads in demand response: A systems and control perspective},'' in
  \emph{{50th IEEE Conference on Decision and Control and European Control
  Conference (CDC-ECC)}}, Dec 2011, pp. 15--20.

\bibitem{Schuss2010}
Z.~Schuss, \emph{Theory and Applications of Stochastic Processes An Analytical
  Approach}, ser. Applied Mathematical Sciences.\hskip 1em plus 0.5em minus
  0.4em\relax Springer, 2010, vol. 170.

\bibitem{Evans2010}
L.~C. Evans, \emph{Partial Differential Equations}, 2nd~ed.\hskip 1em plus
  0.5em minus 0.4em\relax American Mathematical Society, 2010.

\bibitem{DuChateauZachmann1986}
P.~DuChateau and D.~W. Zachmann, \emph{Schaums outline of theory and problems
  of partial differential equations}.\hskip 1em plus 0.5em minus 0.4em\relax
  MacGraw-Hill, 1986.

\bibitem{Hanson2004}
F.~B. Hanson, \emph{Applied Stochastic Processes and Control for Jump
  Diffusions: Modeling, Analysis, and Computation}.\hskip 1em plus 0.5em minus
  0.4em\relax Philadelphia, PA: SIAM Books, 2004.

\bibitem{Oksendal2003}
B.~{\O}ksendal, \emph{Stochastic Differential Equations}, 6th~ed.\hskip 1em
  plus 0.5em minus 0.4em\relax Springer, 2003.

\bibitem{Lee2013}
J.~M. Lee, \emph{Introduction to Smooth Manifolds}.\hskip 1em plus 0.5em minus
  0.4em\relax Springer, 2013.

\bibitem{PaccagnanKamgarpourLygeros2015}
D.~Paccagnan, M.~Kamgarpour, and J.~Lygeros, ``On the range of feasible power
  trajectories for a population of thermostatically controlled loads,'' in
  \emph{54th IEEE Conference on Decision and Control (CDC)}, Dec. 2015, pp.
  5883--5888.

\bibitem{Totu2017}
L.~C. Totu, R.~Wisniewski, and J.~Leth, ``Demand response of {TCL} population
  using switching-rate actuation,'' \emph{IEEE Transactions on Control Systems
  Technology}, vol.~25, no.~5, pp. 1537--1551, Sept. 2017.

\bibitem{Li2014}
S.~Li, W.~Zhang, J.~Lian, and K.~Kalsi, ``On market-based coordination of
  thermostatically controlled loads with user preference,'' in \emph{53rd IEEE
  Conference on Decision and Control (CDC)}, Dec. 2014, pp. 2474--2480.

\bibitem{GridLAB}
\BIBentryALTinterwordspacing
{GridLAB-D} residential module user's guild. [Online]. Available:
  \url{http://sourceforge.net/apps/mediawiki/gridlab-d/index.php?title=
  Residential_module_user%27s_guide}
\BIBentrySTDinterwordspacing

\bibitem{LeVeque2002}
R.~J. LeVeque, \emph{Finite Volume Methods for Hyperbolic Problems}, ser.
  Cambridge Texts in Applied Mathematics.\hskip 1em plus 0.5em minus
  0.4em\relax Cambridge University Press, 2002.

\bibitem{Karatzas1991}
I.~Karatzas and S.~E. Shreve, \emph{Brownian Motion and Stochastic Calculus},
  2nd~ed.\hskip 1em plus 0.5em minus 0.4em\relax Springer, 1991.

\bibitem{Elliott1976}
R.~J. Elliott, ``Stochastic integrals for martingales of a jump process with
  partially accessible jump times,'' \emph{Zeitschrift f{\"{u}}r
  Wahrscheinlichkeitstheorie und Verwandte Gebiete}, vol.~36, no.~3, pp.
  213--226, 1976.

\bibitem{Applebaum2009}
D.~Applebaum, \emph{L{\'{e}}vy Processes and Stochastic Calculus}.\hskip 1em
  plus 0.5em minus 0.4em\relax Cambridge University Press, 2009.

\bibitem{Revuz1999}
D.~Revuz and M.~Yor, \emph{Continuous Martingales and Brownian Motion},
  3rd~ed., ser. Comprehensive Studies in Mathematics.\hskip 1em plus 0.5em
  minus 0.4em\relax Springer, 1999, vol. 293.

\bibitem{Driver2002}
\BIBentryALTinterwordspacing
B.~K. Driver. (2002, Mar.) Surfaces, surface integrals and integration by
  parts. {Lecture Notes on Partial Differential Equations}. [Online].
  Available: \url{www.math.ucsd.edu/~bdriver/231-02-03/Lecture_Notes/pde8.pdf}
\BIBentrySTDinterwordspacing

\bibitem{Durrett2010}
R.~Durrett, \emph{Probability Theory and Examples}, 4th~ed.\hskip 1em plus
  0.5em minus 0.4em\relax Cambridge University Press, 2010.

\end{thebibliography}

\vspace*{-2\baselineskip}
\begin{IEEEbiography}
    [{\includegraphics[width=1in,height=1.25in,clip,keepaspectratio]{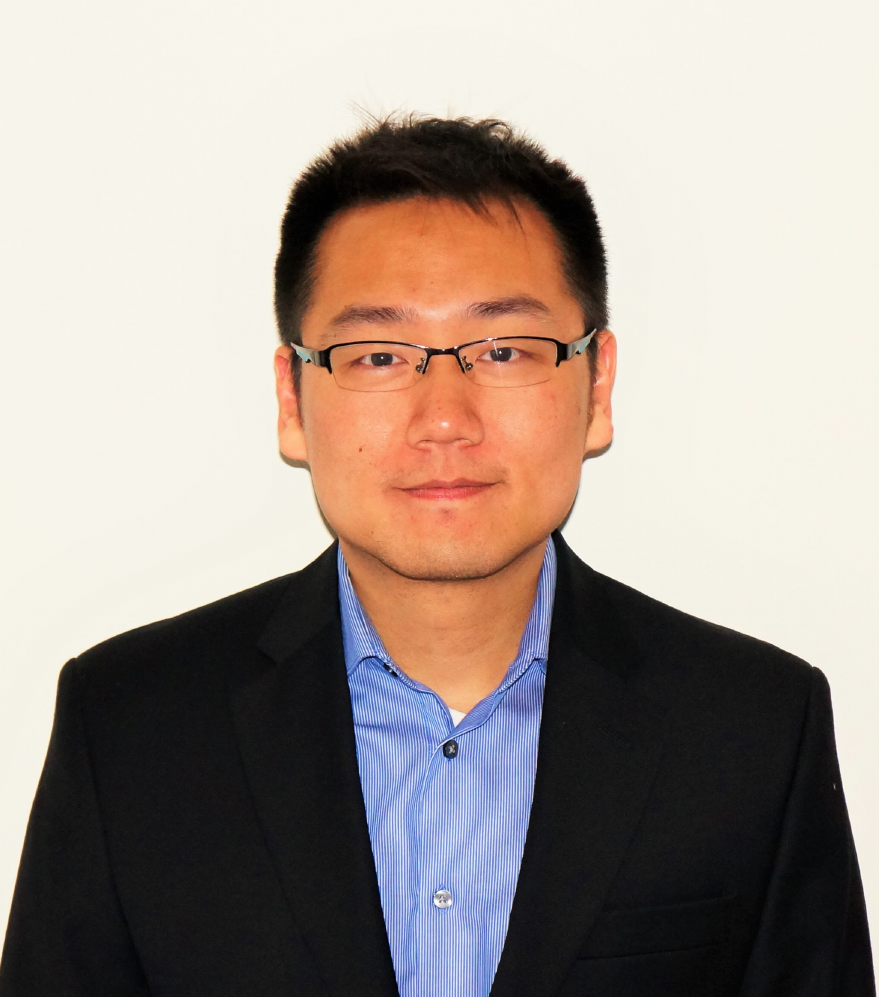}}]{Lin Zhao}
received the B.E. and M.S. degrees in automatic control from the Harbin Institute of Technology, Harbin, China, in 2010 and 2012, respectively. He is currently a Ph.D. student in the Department of Electrical and Computer Engineering, The Ohio State University, Columbus, OH, USA. His current research interests include modeling and control of large-scale complex systems with applications in power systems.
\end{IEEEbiography}

\begin{IEEEbiography}
[{\includegraphics[width=1in,height=1.25in,clip,keepaspectratio]{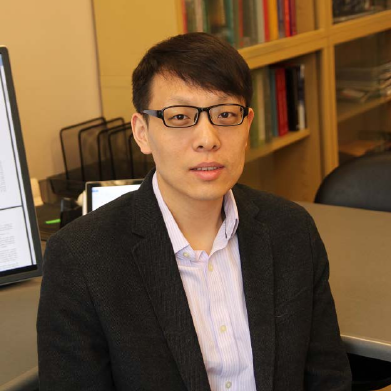}}]{Wei Zhang}
received a B.S. degree in automatic control from the University of Science and Technology of China, Hefei, China, in 2003, and a M.S. degree in statistics and a Ph.D. degree in electrical engineering from Purdue University, West Lafayette, IN, USA, in 2009. Between 2010 and 2011, he was a Postdoctoral Researcher with the Department of Electrical Engineering and Computer Sciences, University of California, Berkeley, CA, USA. He is currently an Associate Professor in the Department of Electrical and Computer Engineering, Ohio State University, Columbus, OH, USA. His research focuses on control and game theory with applications in power systems, robotics, and intelligent transportations.
\end{IEEEbiography}

\end{document}